\PassOptionsToPackage{table}{xcolor}
\documentclass[sigconf]{acmart}
\AtBeginDocument{%
  \providecommand\BibTeX{{%
    \normalfont B\kern-0.5em{\scshape i\kern-0.25em b}\kern-0.8em\TeX}}}

\usepackage{enumitem}
\usepackage{color}

\usepackage{xspace}
\newcommand{\tool}{\textsc{MUSE}\xspace}

\newcommand{\new}[1]{{\color{black}{#1}}}

\newcommand{\mycircle}[1]{{\large 
\textcircled{\small #1}}}

\copyrightyear{2026}
\acmYear{2026}
\setcopyright{cc}
\setcctype{by}
\acmConference[UIST '26]{The 39th Annual ACM Symposium on User Interface Software and Technology}{November 02--05, 2026}{Detroit, MI, USA}
\acmBooktitle{The 39th Annual ACM Symposium on User Interface Software and Technology (UIST '26), November 02--05, 2026, Detroit, MI, USA}
\acmDOI{10.1145/3830398.3830529}
\acmISBN{979-8-4007-2856-3/2026/11}

\begin{document}

\title{MUSE: An Interactive Meta-Agent for Understanding and Steering LLM-powered Data Science Systems}

\author{Wei-Hao Chen}
\email{chen4129@purdue.edu}
\orcid{0009-0003-9108-8473}
\affiliation{%
  \institution{Purdue University}
  \city{West Lafayette}
  \state{Indiana}
  \country{USA}
}

\author{Weixi Tong}
\email{tong172@purdue.edu}
\orcid{0009-0001-2318-3191}
\affiliation{%
  \institution{Purdue University}
  \city{West Lafayette}
  \state{Indiana}
  \country{USA}
}

\author{Yuan Tian}
\email{tian211@purdue.edu}
\orcid{0000-0002-2200-5713}
\affiliation{%
  \institution{Purdue University}
  \city{West Lafayette}
  \state{Indiana}
  \country{USA}
}

\author{Chenglong Wang}
\email{chenwang@microsoft.com}
\orcid{}
\affiliation{%
  \institution{Microsoft Research}
  \city{Redmond}
  \state{Washington}
  \country{USA}
}

\author{Tianyi Zhang}
\email{tianyi@purdue.edu}
\orcid{0000-0002-5468-9347}
\affiliation{%
  \institution{Purdue University}
  \city{West Lafayette}
  \state{Indiana}
  \country{USA}
  \postcode{43017-6221}
}


\begin{abstract}
Recent advances in large language models have enabled a new class of agentic data science systems that allow users to complete complex data science workflows through natural language. Although these systems can significantly reduce manual effort, it remains difficult to diagnose their behavior and steer the reasoning process when failures or unexpected outputs occur. We present \tool, an interactive meta-agent that enhances user understanding and control of agentic data science systems by (1) dynamically restructuring low-level execution traces into multiple semantic levels that support navigation from high-level overviews to low-level implementation details; (2) enabling users to reference specific workflow steps in context to ask grounded questions, provide feedback, and revise problematic steps without manually locating relevant execution history; and (3) supporting mixed-initiative steering by surfacing suspicious steps for inspection, scaffolding the repair process, and translating user repair intent into contextualized instructions for the underlying agent. In a between-subjects study ($n=15$), \tool improved task efficiency and increased users' confidence in understanding and steering agentic data science workflows.
\end{abstract}




\begin{teaserfigure}
    \centering
  \includegraphics[
    width=.85\textwidth,
    trim=1.8cm 4.3cm 3.8cm 3cm,
    clip
  ]{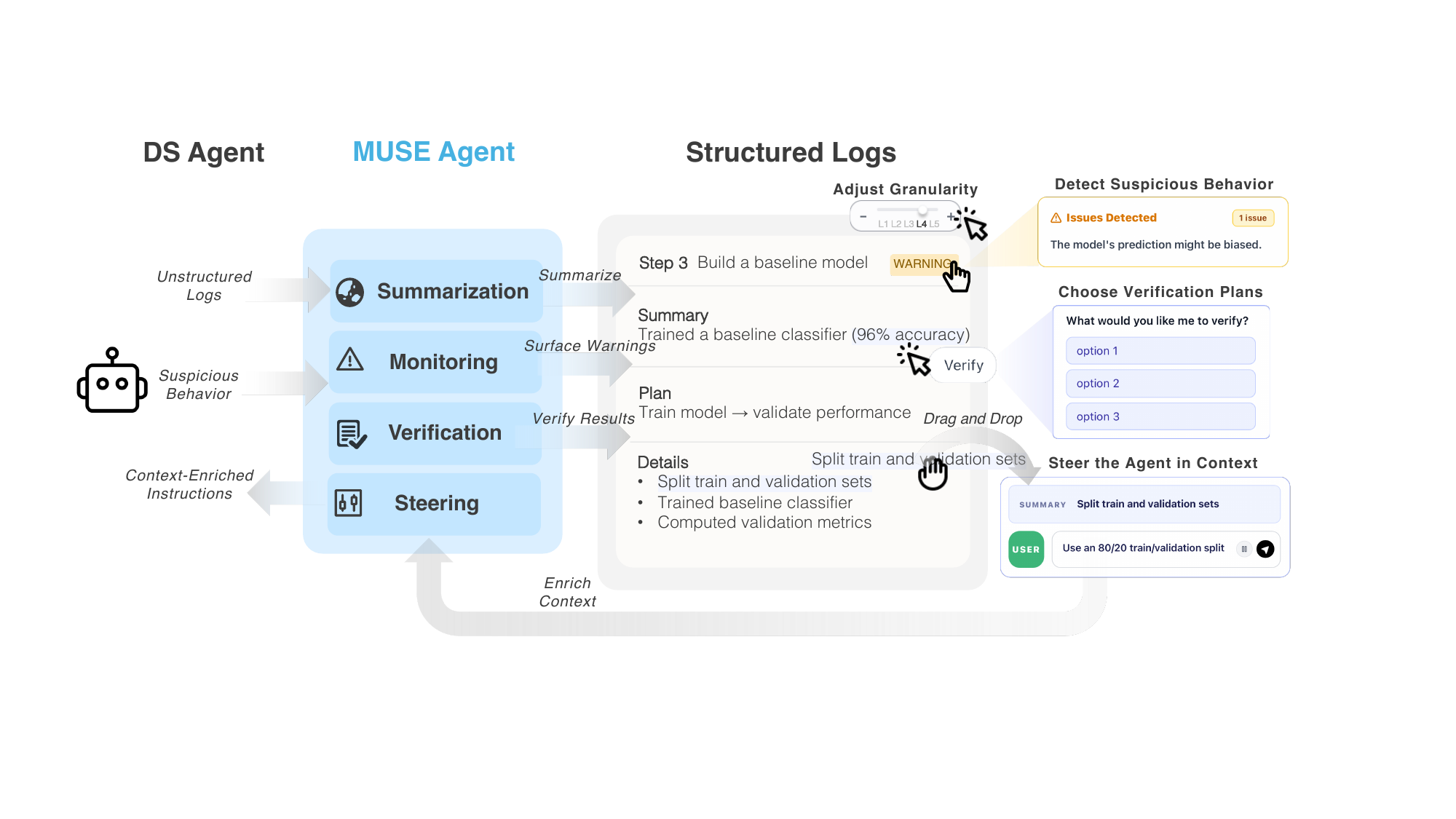}
  \vspace{-10pt}
\caption{
\textbf{Left:} An underlying data science agent that produces unstructured execution logs, including code edits, tool calls, and runtime outputs.
\textbf{Middle:} MUSE serves as a meta-agent that transforms these raw logs into structured logs with multiple semantic levels and monitors the process to detect suspicious behavior and surface warnings at runtime.
\textbf{Right:} Users interact with the structured logs to adjust semantic granularity, inspect surfaced issues, specify what to verify through scaffolded, in-context verification plans, and steer the agent in context by providing feedback and revising specific workflow steps.}
  \Description{}
  \label{fig:teaser}
\end{teaserfigure}


\maketitle

\section{Introduction}
Data science (DS) is inherently complex and iterative~\cite{kery2018story}. In practice, data science workers often spend substantial time inspecting data, writing one-off scripts, cleaning messy datasets, and stitching together heterogeneous data sources~\cite{wang2021autods, chattopadhyay2020s, muller2019data, kery2017exploring}. They often ``\textit{have a conversation}'' with their data, iteratively probing patterns, interpreting intermediate results, and refining their analytical questions over time~\cite{muller2019data}. As a result, data analysis involves not only carrying out analytical steps but also revisiting earlier decisions and adjusting the workflow as new findings emerge.

Recent advances in large language models (LLMs)~\cite{achiam2023gpt, team2024gemini, touvron2023llama} have enabled a new class of agentic systems~\cite{zhang2025deepanalyze, nam2025mle} that orchestrate specialized agents to perform tasks such as data cleaning, feature engineering, and visualization.  These systems enable users to complete data science tasks by specifying their needs in a few natural language prompts, making sophisticated tasks more accessible to users with limited expertise. For example, a sales manager may identify factors associated with customer churn without manually writing complex analytical code.

Although such agentic systems can greatly reduce human effort, they provide limited transparency for users to steer the reasoning process.
For example, users need to sift through large volumes of low-level execution traces to determine what went wrong. While prior systems~\cite{googleadk,epperson2025interactive, sheng2026dills} support monitoring and inspection of agentic systems, they are still designed for professional \textit{agent developers}. They expose low-level observability, such as agent messages, internal states, and tool invocations. For non-expert users, such information can be overwhelming.
Furthermore, existing tools provide limited support for correcting erroneous steps. Users need to manually describe repairs in free-form language, making it difficult for users with limited prompting expertise~\cite{zamfirescu2023johnny}. 



To better understand challenges in AI-assisted data analysis, we conducted a formative study with 7 participants who regularly use AI-assisted tools for data analysis. The study showed that the main difficulty was not simply seeing low-level system logs, but judging whether the overall workflow made sense. Participants reported that debugging system behavior was difficult because raw logs provided little structure for assessing system progress. They also found repair difficult because fixing a problem required tracing low-level implementation details and translating high-level analytical intentions into concrete changes to generated code or prompts.

To address these challenges, we propose \tool, an interactive \textit{meta-agent} that mediates between users and underlying agent systems. \tool restructures low-level execution logs into multiple semantic levels, ranging from high-level workflow overviews to low-level implementation traces. This representation allows users to inspect system behavior at a level of detail that matches their needs and understanding.
\tool also enables users to drag workflow steps into the chat interface to provide feedback and revise problematic steps in context.
It enriches these requests with relevant workflow context and translates them into contextualized instructions, without requiring users to manually locate or restate that context. In addition, \tool detects suspicious system behavior and surfaces potential errors to help users identify problematic steps. When suspicious steps are surfaced, users can revise them through scaffolded repair, allowing them to specify how a step should be changed and steer the workflow. Finally, \tool scaffolds the verification process by suggesting verification plans, discussing what intermediate results to verify, and generating targeted verification scripts grounded in the current workflow.


We conducted a between-subjects user study with 15 participants to evaluate the usability and efficiency of {\tool}. Participants using {\tool} completed the  tasks in 17 minutes on average, reducing task completion time by 35\% and 9\% compared with the other two conditions. In the post-task survey, participants also reported feeling more confident about the results when using {\tool}.

\section{Related Work}

\subsection{Interactive Data Science Systems} 
Data science workflows are inherently exploratory and iterative. In his seminal work on Exploratory Data Analysis (EDA)~\cite{tukey1977exploratory}, Tukey emphasized the open-ended, discovery-driven nature of working with data. Such workflows typically span multiple stages including data cleaning, data exploration, modeling, and evaluation. HCI researchers have developed many interactive tools to support these stages. For example, one of the earliest data exploration systems was Ahlberg and Shneiderman's dynamic queries~\cite{ahlberg1994visual}, which enabled users to filter datasets through real-time slider manipulations with immediate feedback. Later, Polaris~\cite{stolte2002polaris} grounded visualization specification in table algebra through a shelf-based drag-and-drop interface, forming the foundation of Tableau. A parallel line of work focused on data wrangling. SMARTedit~\cite{lau2001learning} applied programming by demonstration to automate text editing tasks, while SWYN~\cite{blackwell2001swyn} enabled users to select text examples and preview the effects of induced transformation scripts. For structured tabular data, Wrangler~\cite{kandel2011wrangler} pioneered mixed-initiative transformation recommendation based on user demonstrations and Wrex~\cite{drosos2020wrex} extended this interaction paradigm to computational notebooks.



The rise of large language models (LLMs)~\cite{team2024gemini, achiam2023gpt, touvron2023llama} marked a further shift in how users interact with data science systems, moving from specifying \textit{what to do} to expressing \textit{what results they want}, a paradigm Nielsen terms \textit{intent-based outcome specification}~\cite{nielsen2023ai}. LLM-based systems such as Data Formulator~\cite{wang2023data, wang2025data} allowed analysts to express analytical goals in natural language and iteratively receive LLM-generated charts and transformations. Dango~\cite{chen2025dango} leveraged LLMs and clarification questions to support mixed-initiative data wrangling. More recently, multi-agent systems such as MLE-STAR~\cite{nam2025mle} and DeepAnalyze~\cite{zhang2025deepanalyze} orchestrated specialized agents across the full data science pipeline.

Although LLM-powered systems have allowed users to perform complex data science tasks through natural language, the \textit{gulf of evaluation} has widened. Recent studies~\cite{sheng2026dills, felder2025retrace} have found that modern LLMs produce verbose, low-level execution traces that are difficult for users to interpret. \new{Kazemitabaar et al.~\cite{kazemitabaar2024improving} and WaitGPT~\cite{xie2024waitgpt} help users verify and refine the DS code generated by an LLM, instead of the agent behavior trajectory. Specifically, Kazemitabaar et al.~\cite{kazemitabaar2024improving} allow users to edit task execution plans and assumptions to refine the code, while WaitGPT~\cite{xie2024waitgpt} visualizes generated code in a graph, allows users to inspect intermediate data, and directly manipulates operation nodes in the graph. By contrast, \tool supports more agentic verification and refinement by generating verification plans and scripts for users to choose from (Figure~\ref{fig:UI}\mycircle{E} and Figure~\ref{fig:verification}), proactively detecting potential errors, and recommending revisions (Figure~\ref{fig:warning}). Furthermore, both prior systems focus on data queries and data analysis tasks, whereas \tool supports more sophisticated end-to-end DS workflows that include data cleaning, feature engineering, and model training.
}

\subsection{Debugging Agentic Systems}

To support development and debugging, many agent frameworks~\cite{ClaudeSDK, googleadk, openaiagentssdk, crewai, autogen, langgraph, langsmith, phoenix, langfuse, microsoftdevui, agentops} provide instrumentation for recording agent events. Tools such as LangSmith~\cite{langsmith} and Phoenix~\cite{phoenix} offer hierarchical trace views, while Langfuse~\cite{langfuse} and Microsoft DevUI~\cite{microsoftdevui} provide graph-based workflow summaries. AgentOps~\cite{agentops} and LangTrace~\cite{langtrace} further support observability dashboards for monitoring. While these tools make it easier to inspect system behavior, they primarily emphasize low-level agent traces, events, and runtime states, and therefore are better suited to users with expertise in developing agentic systems. 

Some recent systems~\cite{epperson2025interactive, sheng2026dills} have begun to support interactive debugging. AGDebugger~\cite{epperson2025interactive} enables users to edit message histories, reset execution states, and re-run workflows from intermediate points. However, such interaction still places the burden on users to write effective prompts, which can be challenging for non-AI-experts~\cite{zamfirescu2023johnny}. While \new{DiLLS~\cite{sheng2026dills} also supports interactive diagnosis through layered summaries of agent behavior, \tool differs from it in three ways. First, the structured summary generated by DiLLS is largely static and users can't act on it. By contrast, \tool allows users to ask clarification questions about a specific step, validate its correctness, and provide targeted feedback. Second, DiLLS relies on users to interpret the summary and identify suspicious behavior, while \tool is more mixed-initiative. Specifically, \tool proactively surfaces suspicious behavior as warnings, provides explanations, and suggests revisions. Third, DiLLS performs post-mortem analysis and generates a summary only after the agent has failed. By contrast, \tool continuously reads agent logs and updates the summary, giving users a live view of the agent's behavior. Furthermore, there is minimal wait time for users since the summary view is updated continuously as the agent makes progress, and users can intervene and provide immediate feedback to the agent at any time.
}

Despite these advances, existing tools are largely designed for \textit{agent developers} who build, debug, and optimize agent systems. They center on execution-level inspection, message-level intervention, or agent-centric explanation. As a result, users often need to translate low-level traces or per-agent activities into their own understanding of the overall data-science workflow and manually formulate revisions based on scattered context. In contrast, \tool restructures low-level execution logs into multiple levels of data-science semantics, ranging from high-level workflow overviews to low-level implementation traces, allowing users to inspect system behavior at a level that matches their understanding. \tool also highlights suspicious steps with localized warnings, scaffolds verification by helping users clarify what to verify and generating targeted verification scripts grounded in the current workflow.



\section{Formative Study}

To understand the challenges of performing DS tasks with current LLM-powered tools, we conducted a formative study with 7 data scientists with experience using AI coding tools. Participants' backgrounds are reported in Table~\ref{tab:formative_participants}. We describe our study procedure in Appendix~\ref{appendix:formativeProcedure}. Based on our observations and interviews, we identified 4 major user needs in Section~\ref{formative:usersneeds}. Finally, we discuss the resulting design rationale in Section~\ref{formative:designrationale}.

\begin{table}[h]
\centering
\setlength{\tabcolsep}{4pt}
\resizebox{\columnwidth}{!}{%
\begin{tabular}{c l l l l}
\toprule
\textbf{PID} & \textbf{Gender} & \textbf{Background} & \textbf{AI Tool Usage} & \textbf{Degree} \\
\midrule
P1 & Male   & Engineering & Cursor & PhD \\
P2 & Male   & Engineering & OpenCode & PhD \\
P3 & Male   & Engineering & Cursor & PhD \\
P4 & Male   & Engineering & Claude Code, Cursor & PhD \\
P5 & Male   & Engineering & Claude Code, Codex & PhD \\
P6 & Female & Statistics  & ChatGPT, Gemini & PhD \\
P7 & Female & Design      & ChatGPT, Claude & PhD \\
\bottomrule
\end{tabular}%
}
\caption{\new{Demographics and background of the formative study participants.}}
\label{tab:formative_participants}
\vspace{-20pt}
\end{table}

\subsection{User Needs}
\label{formative:usersneeds}
\noindent\textbf{N1:\textit{ Understand system behavior in the DS workflow (P1, P2, P3, P6, and P7).}}  Existing systems often expose execution details without clearly conveying the overall DS workflow. Because these systems only surfaced low-level traces, participants struggled to map such outputs onto their own mental model of how a DS workflow should unfold. For example, P3 noted that the tool ``\textit{outputs too many code pieces without sufficient explanations, making it hard to tell whether the system was cleaning data, training a model, or carrying out some other step.}'' P2 noted that not all logs needed to be visible, as long as they could still understand the system's overall progress. These responses suggest a need for better representations rather than simply exposing implementation traces.

\noindent\textbf{N2:\textit{ Quickly identify which steps may be worth reviewing (P1, P3, P4, and P7).}} They often found it hard to tell where problems had occurred or what had failed. P4 noted that ``\textit{the tool outputs too many responses that make me feel overwhelmed and make it difficult to locate the error,}'' and P3 similarly described the output as ``\textit{very hard to verify.}'' We also observed that some participants did not inspect each intermediate trace and instead relied on the tool unless an output seemed incorrect (P1, P2, and P4). P7 further suggested that the interface should highlight ``\textit{the most important part related to the failure.}'' These responses suggest that users need help quickly spotting which workflow steps may be worth reviewing.

\noindent\textbf{N3:\textit{ Specify repair strategy precisely in context (P2, P4, P6, and P7).}}
Participants often needed to communicate not only \textit{what} should change, but also \textit{where} in the workflow the change should be applied and \textit{how} that part should be revised. They had to manually read through logs, find the relevant parts, and copy and paste them into the chat to specify their intent. Even with this effort, relevant details could still be missed, making it difficult to steer the system toward the intended part of the workflow. P2 suggested ``\textit{an interactive UI that could anchor feedback to a specific part of the generated workflow.}'' These responses suggest a need for mechanisms that support more contextual and precise specification.

 \noindent\textbf{N4:\textit{ Mixed-initiative control during execution (P3, P4, P6, and P7).}} Participants wanted to interrupt, redirect, or verify the process while it was running, rather than waiting until the workflow finished. P3 and P4 noted that they could not interact with the model during long-running execution. For example, training a model could take a long time, and the system could get stuck only showing \textit{``spinning.''} P7 similarly noted that they needed to \textit{``wait until all the work was done''} before checking the results. Participants also wanted better ways to \textit{participate} in the decision-making process. P3 and P7 suggested that the tool should consult with them before taking actions that could lead to unintended results. They also wanted a better way to verify their hypotheses. These responses suggest a need for mixed-initiative interaction that allows users to steer or repair the workflow while it is running.
\subsection{Design Rationale}
\label{formative:designrationale}

Prior tools~\cite{epperson2025interactive, langfuse, langgraph, langsmith} typically surface raw agent messages, tool calls, or code traces directly. However, such representations still leave users facing the \textit{coordination barrier}~\cite{ko2004six}. To support \textbf{N1}, \tool translates unstructured logs into a structured representation with five levels of semantics. \tool provides representations ranging from high-level summaries to detailed execution evidence. Users can navigate these levels using semantic sliders, starting from an overview and drilling down into lower-level details when needed. This design is inspired by the information-seeking mantra~\cite{shneiderman2003eyes}, which emphasizes \textit{``overview first, zoom and filter, and details on demand.''} In addition, \tool explains system behavior in terms of DS workflow semantics rather than individual agent behavior, since agent-level explanations still require users to reason about the underlying coordination logic.

To support \textbf{N2}, \tool monitors the underlying DS agent at runtime, operating in parallel to check for suspicious system actions. \tool then surfaces these issues as warnings attached to the corresponding stages, providing \textit{localized cues} that help users quickly identify which steps may be worth reviewing. This design is inspired by the notion of \textit{information scent} from information foraging theory~\cite{pirolli1995information, chi2001using}, directing users' attention to where useful diagnostic information is most likely to be found. Once a warning is surfaced, \tool allows users to inspect it and further scaffolds the repair process to help users correct erroneous workflow steps.

To support \textbf{N3}, \tool allows users to drag and drop a workflow step into the chatbox as the target of their request. Each workflow step across semantic levels is linked to its underlying code fragments and execution logs, allowing \tool to resolve the selected step to the relevant evidence, enrich the request with workflow context, and augment the user’s prompt with the information needed to make the request actionable. This design has two key benefits. First, users can express repairs at the level of workflow semantics rather than low-level implementation details. Second, \tool lowers the prompting burden by sparing users from manually locating logs, tracing code fragments, or restating execution history—a process that can be challenging for non-AI-expert users~\cite{zamfirescu2023johnny}.

To support \textbf{N4}, \tool enables mixed-initiative control~\cite{horvitz1999principles} during execution. Because \tool operates in parallel with the underlying agent, it can inspect execution traces without waiting for the current step to finish. This allows users to ask questions and understand the system during long-running steps without stopping the underlying agent. To help users build trust in the agent, \tool further scaffolds the processes of verification and repair. Specifically, for repair, \tool guides users in specifying how a problematic step should be changed by presenting actionable repair options grounded in the current workflow context. For verification, \tool asks follow-up questions to clarify what the user wants to check and then generates a targeted verification script grounded in the relevant artifacts. In this way, \tool turns both repair and verification into scaffolded, in-context interactions that reduce the burden of deciding \textit{what} to change or verify, and \textit{how} to do so.

\begin{figure*}[ht!]
    \centering
    \includegraphics[
        width=0.9\textwidth,
        trim=1cm 6.1cm 1cm 5.62cm,
        clip
    ]{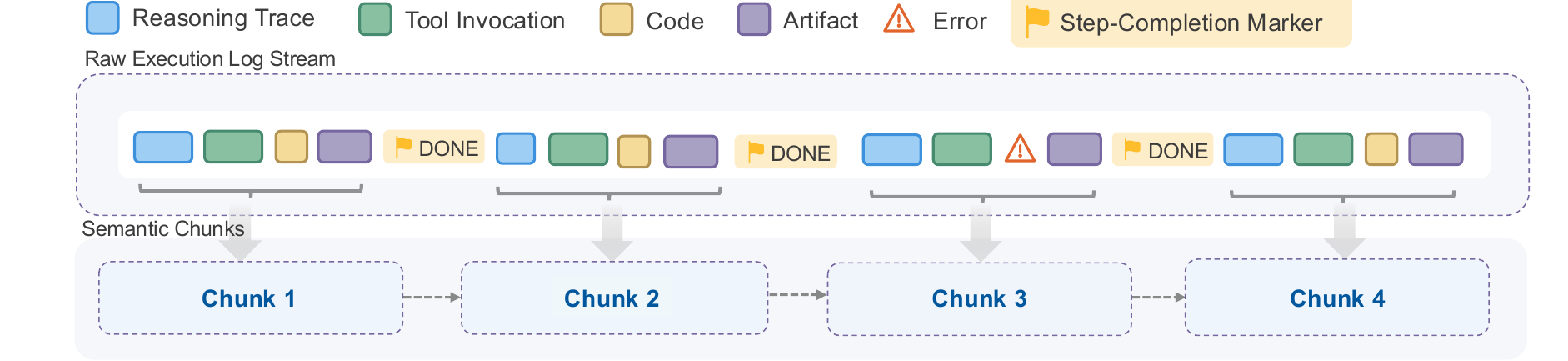}
\caption{\new{MUSE dynamically converts the underlying DS agent's raw execution log stream---including reasoning traces, tool invocations, code, artifacts, and errors---into semantic chunks using explicit step-completion markers.}}
    \label{fig:semantic_chunk}
\end{figure*}

\section{System Design}
\label{system_design}

As shown in Figure~\ref{fig:teaser}, \tool is organized into four layers: (1) the \textit{Summarization Layer}, which transforms raw execution logs into structured representations; (2) the \textit{Monitoring Layer}, which highlights potentially problematic steps; (3) the \textit{Verification Layer}, which supports in-situ verification of intermediate results through scaffolded verification plans and targeted verification scripts; and (4) the \textit{Steering Layer}, which enables users to provide feedback, revise problematic workflow steps, and steer system behavior. \tool is designed as a meta-agent that acts upon an underlying data science agent that performs common tasks such as data preprocessing, model training, and evaluation. Implementation details of the underlying DS agent are provided in Appendix~\ref{appendix:underlyingDSAgent}.

\subsection{\textbf{Summarization Layer}}

This layer dynamically converts unstructured logs into meaningful explanations through three stages: (1) capturing agent system execution traces, (2) grouping them into semantic chunks, and (3) transforming each chunk into a multi-level semantic representation.

\subsubsection{\textbf{Agent System Execution Trace}}
\begin{figure*}[ht!]
    \centering
    \includegraphics[
        width=0.9\textwidth,
        trim=0.2cm 0.25cm 0.4cm 0.2cm,
        clip
    ]{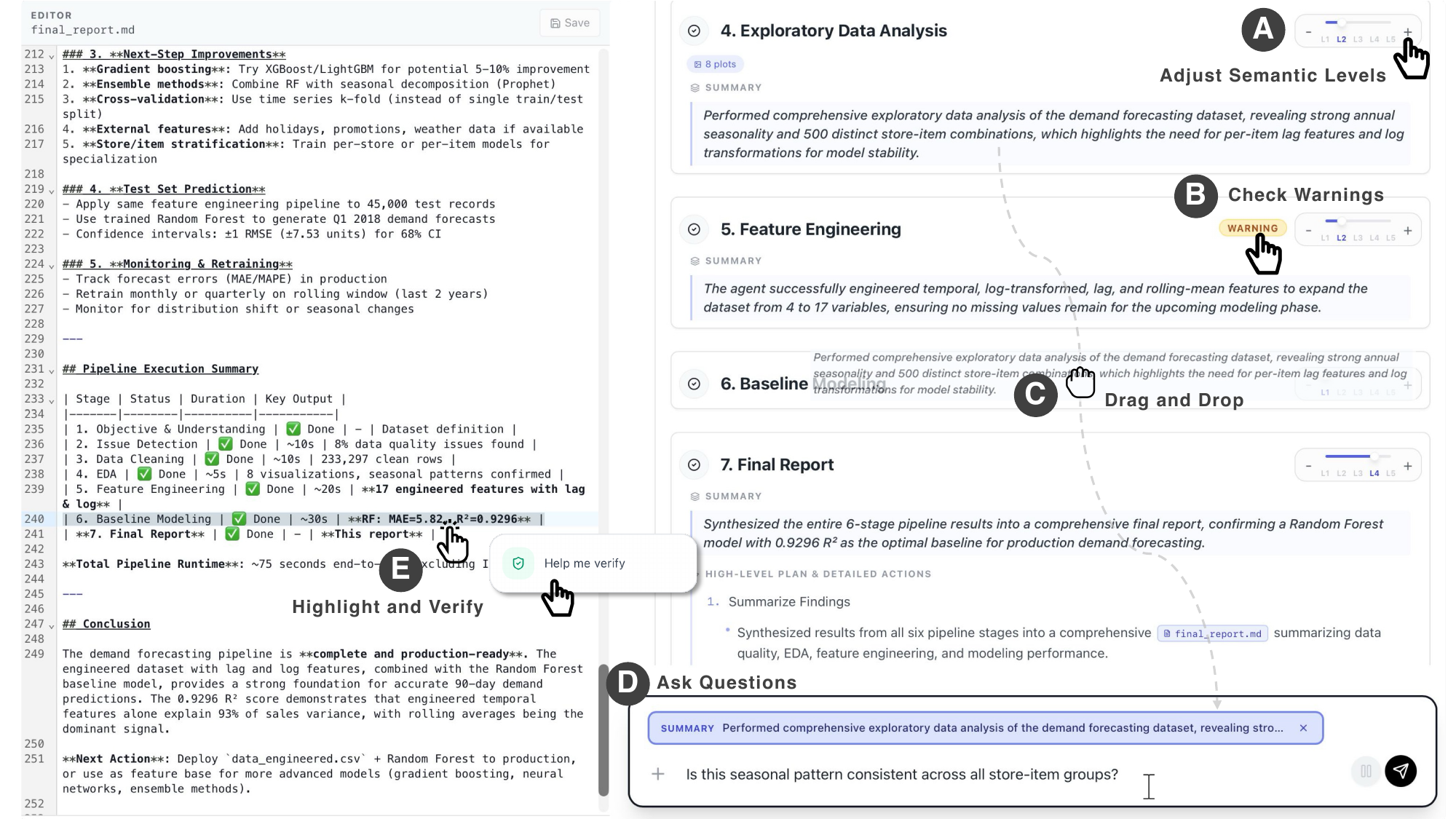}
    \vspace{-10pt}
    \caption{
    \tool{} provides an interactive interface for understanding and steering agentic data science systems.
    \mycircle{A} Users can adjust semantic levels (L1–L5) to navigate between high-level summaries and detailed execution traces.
    \mycircle{B} MUSE surfaces localized warnings on workflow steps to highlight potentially problematic behaviors.
    \mycircle{C} Users can drag and drop a step into the chatbox.
    \mycircle{D} Users ask follow-up questions in context without needing to reconstruct execution details.
    \mycircle{E} Users can highlight outputs (e.g., metrics) to trigger an in-situ verification process.
    }
    \label{fig:UI}
\end{figure*}
\tool instruments the agent runtime and records a stream
of execution events. These events are emitted whenever the underlying
agent produces intermediate outputs, invokes tools, or executes code. During execution, the backend
writes these events to a persistent local trace log file, which
serves as the primary observation channel for \tool. Specifically, \tool captures the following categories of execution
logs:

\begin{enumerate}
\renewcommand{\labelenumi}{\arabic{enumi}.}
\item \textbf{\lbrack \textsc{REASONING TRACE}\rbrack}: internal reasoning or planning logs produced by the underlying DS agent during execution.
\item \textbf{\lbrack \textsc{TOOL INVOCATION}\rbrack}: tool calls, including tool names, input arguments, and outputs.
\item \textbf{\lbrack \textsc{CODE}\rbrack}: code snippets and system commands generated by the underlying DS agent.
\item \textbf{\lbrack \textsc{SIGNAL}\rbrack}: signals and status indicators, including warnings, errors, and step-completion markers.
\item \textbf{\lbrack \textsc{ARTIFACT}\rbrack}: intermediate outputs, such as dataset summaries and model evaluation results.
\end{enumerate}

\new{
\subsubsection{\textbf{Semantic Chunk}}
As shown in Figure~\ref{fig:semantic_chunk}, \tool dynamically organizes the agent system's execution logs into a sequence of \textit{semantic chunks}. \tool segments the logs into stage-level units. The underlying DS agent executes a sequence of stages such as data cleaning or feature engineering; the DS agent emits a \textit{step-completion marker} immediately after it finishes a stage. The chunk boundaries are determined by the positions of these markers in the log stream. This allows \tool to segment the trace according to the stages the agent has actually completed, rather than relying on an LLM to infer boundaries from free-form logs. As soon as a chunk boundary is detected, the chunk is translated into a multi-level semantic representation, keeping the interface synchronized with the agent's progress.
}


\subsubsection{\textbf{Multi-Level Semantic Representation}}

\new{Each semantic chunk is transformed into a semantic representation with five progressively revealed levels of abstraction, supporting \textit{overview-to-detail} navigation.} \textbf{L1} presents the title of the DS stage. \textbf{L2} adds a natural-language summary of what was accomplished. \textbf{L3} presents an ordered list of sub-steps. \textbf{L4} expands this list with detailed actions, such as generated files, executed commands, and selected parameters. Finally, \textbf{L5} exposes the traces, including code, commands, and tool outputs. \tool generates this representation in real time using a prompt (Table~\ref{tab:translation_prompt}); Figure~\ref{fig:data_cleaning_card} shows an example.


\subsection{\textbf{Monitoring Layer}}
The Monitoring Layer detects suspicious behaviors in each semantic chunk and surfaces localized warnings for users to act on.

\subsubsection{\textbf{Warning Detection}}

\tool applies a set of warning detection heuristics to identify semantic chunks that may require user attention, capturing suspicious behaviors such as reasoning--outcome mismatches, incorrect data processing, execution errors, and other forms of agent misbehavior. These heuristics were developed from 20 data science tasks in MLE-Bench~\cite{chan2024mle-bench} and informed by prior taxonomies of agent misbehavior~\cite{cemri2025multi}. Table~\ref{tab:detailed_warning_heuristics} summarizes the heuristics, and Table~\ref{tab:warning_prompt} presents the full prompt.

\subsubsection{\textbf{Localized Warnings and Follow-up Actions}}
When a semantic chunk is identified as suspicious, \tool presents a \textit{localized warning} badge directly on the corresponding step card, as shown in Figure~\ref{fig:warning}. Hovering over the badge opens an inline warning panel that explains the detected issue in natural language and offers follow-up actions. Compared to the warnings in DiLLS~\cite{sheng2026dills}, which mainly serve as static cues for inspection, \tool supports both warning \textit{understanding} and \textit{follow-up repair}. Users can click the ``Why this warning?'' button to inspect why the step was flagged. When users request an explanation, \tool presents a grounded description of the detected issue based on the warning diagnosis with evidence from the corresponding semantic chunk, such as relevant file names, as shown in Figure~\ref{fig:warning}\mycircle{2}. This allows users to understand what triggered the warning and which part of the workflow it refers to. If users decide to revise the workflow, \tool scaffolds the repair process by helping them determine how the issue should be fixed, as shown in Figure~\ref{fig:warning}\mycircle{3}.

\subsection{\textbf{Verification Layer}}
The Verification Layer helps users verify whether intermediate results and workflow outputs are correct.

\subsubsection{\textbf{Contextualized Referencing}}

As shown in Figure~\ref{fig:UI}\mycircle{C}, users can drag a workflow step into the chat interface to ask what the step is doing, why it was performed, or whether an intermediate result appears reasonable. \tool treats each drag-and-drop action as an anchor. \tool resolves this anchor against the session context and retrieves relevant reports, scripts, artifacts, and other sandbox files as prompt evidence.
 As a result, \tool answers user questions using grounded evidence rather than the query alone, enabling localized and context-rich interaction without requiring users to manually search logs or restate execution history. Table~\ref{tab:contextualized_referencing_prompt} shows the prompt template used.

\subsubsection{\textbf{Scaffolded Verification}}
\tool supports verification in the sandbox environment and can verify results independently without instructing the DS agent. Users can highlight an output and right-click to invoke \textit{Help me verify}, as shown in Figure~\ref{fig:UI}\mycircle{E}. Rather than requiring users to manually decide what to check, \tool combines the selected output with its associated file and session context, then generates verification plans for the user to choose from, as shown in Figure~\ref{fig:verification}\mycircle{1}. After a plan is selected, \tool inspects the reports, scripts, metrics, and other files, and, when needed, generates  verification scripts inside the session sandbox without modifying pipeline outputs, sparing users from manually writing or running these checks themselves, as shown in Figure~\ref{fig:verification}\mycircle{2}. Table~\ref{tab:verification_prompt} and Table~\ref{tab:verification_execution_prompt} show the prompt templates.

\subsection{\textbf{Steering Layer}}

The Steering Layer enables users to repair the \textit{ongoing} workflow.

\subsubsection{\textbf{Scaffolded Repair}}
\begin{figure}[ht!]
    \centering
    
    \includegraphics[width=0.72\linewidth,
            trim=0.2cm 0.1cm 0.2cm 0.1cm,
        clip]{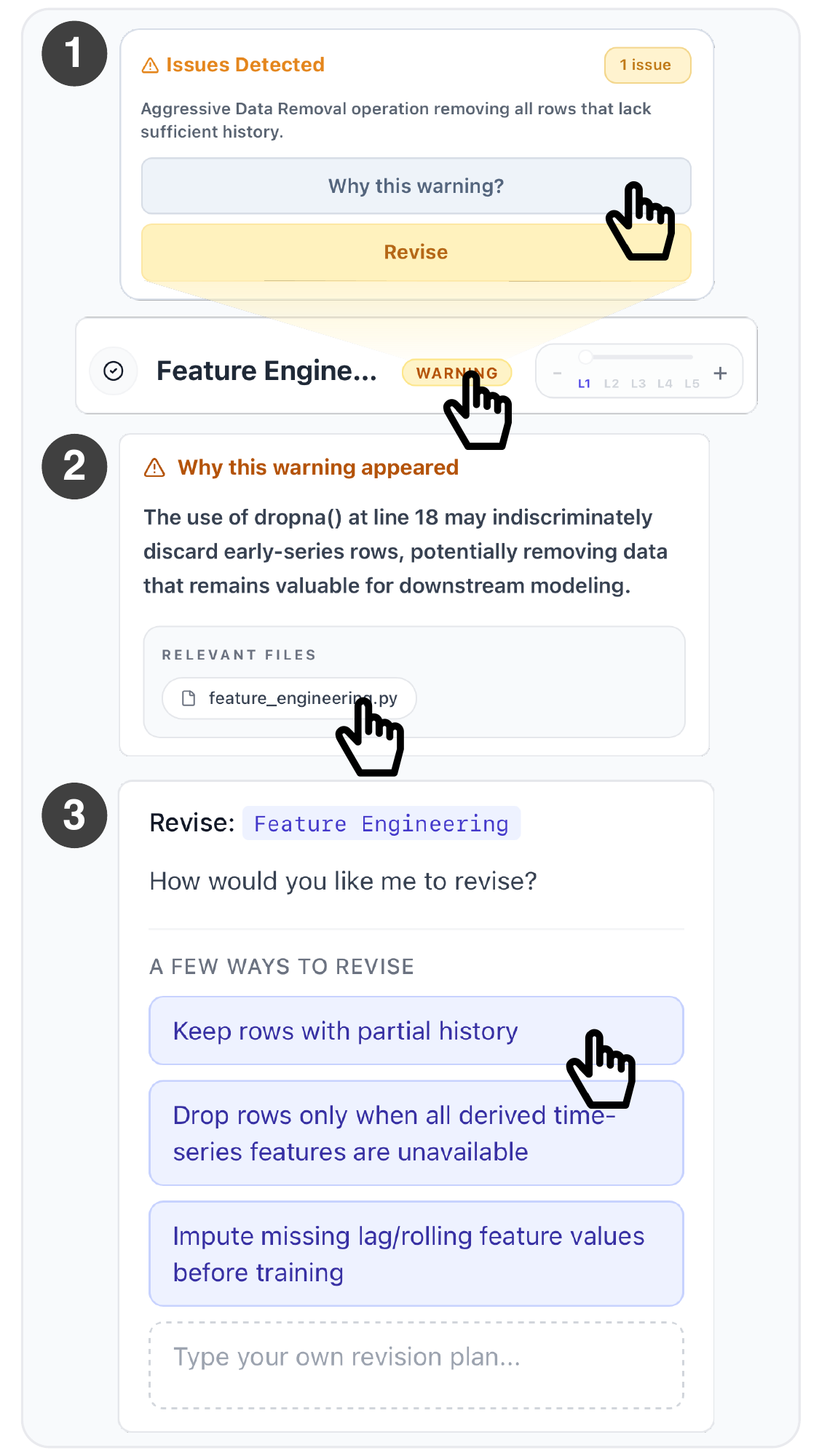}
    \vspace{-10pt}
    \caption{\tool surfaces warnings, allowing users to inspect issues, request explanations, or revise the workflow.}
    
    \label{fig:warning}
\vspace{-10pt}
\end{figure}

Users can invoke \textit{Revise}, as shown in Figure~\ref{fig:warning}\mycircle{1}. Rather than requiring users to manually diagnose the issue, the repair interface is scaffolded using the warning explanations and repair suggestions generated by \tool. The warning explanation also references the relevant code or data locations, allowing users to inspect the referenced files directly by clicking the file icon, as shown in Figure~\ref{fig:warning}\mycircle{2}. Users can then choose a suggested repair plan, as shown in Figure~\ref{fig:warning}\mycircle{3}. \tool then automatically combines the repair request together with the affected step and warning context into a contextualized instruction, sparing users from manually assembling this information, and passes it to the underlying DS agent. Table~\ref{tab:revise_handoff_template} shows the prompt template.

\begin{figure}[ht!]
    \centering
    \includegraphics[width=0.69\linewidth,
            trim=0cm 4.7cm 0.1cm 4.2cm,
        clip]{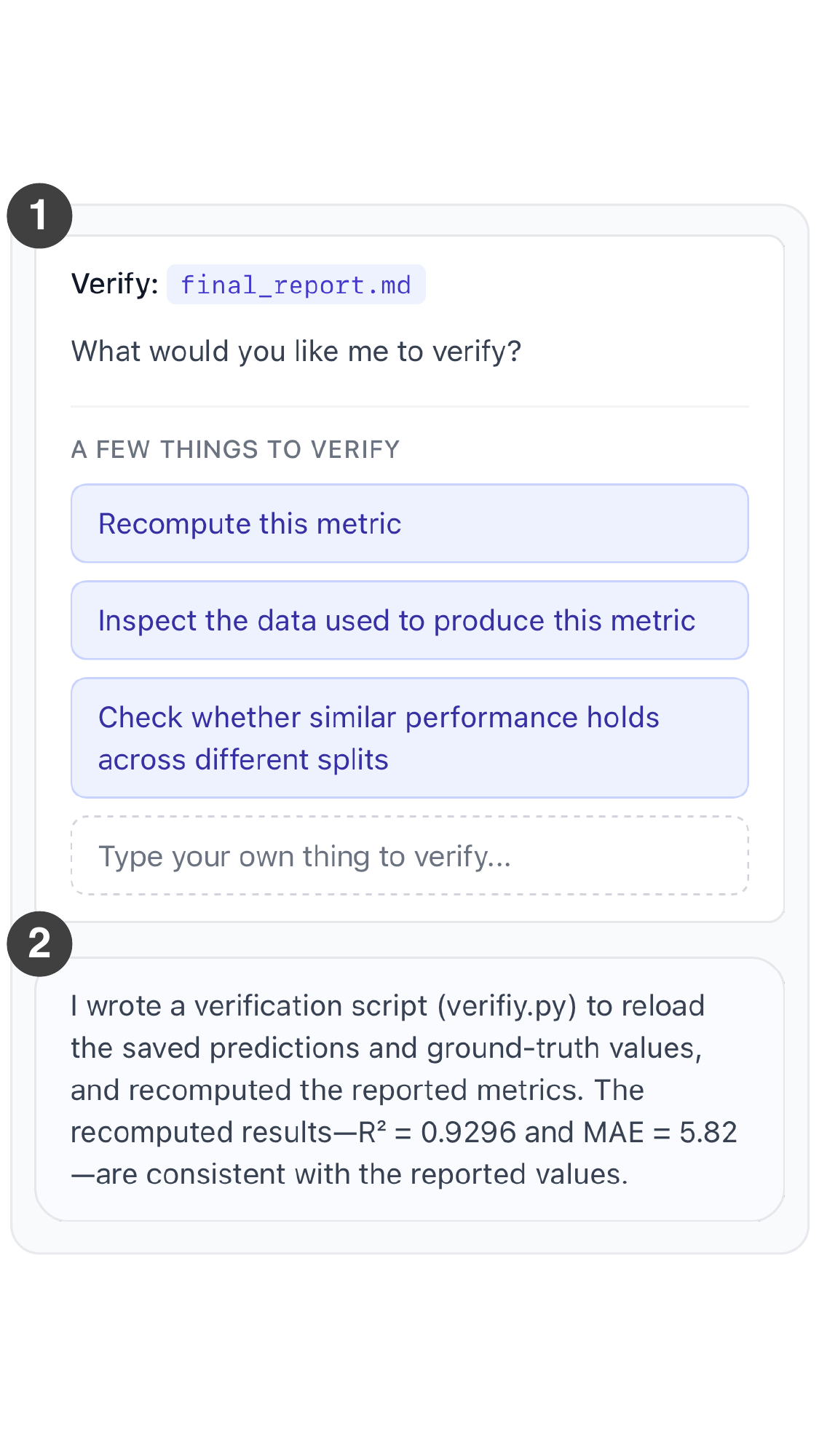}
    \vspace{-10pt}
    \caption{\tool asks follow-up questions and builds a verification script to validate the result.}
    \label{fig:verification}
\end{figure}
\section{USAGE SCENARIO}
\new{
Johnny is a sales manager who wants to forecast future demand using his company's historical sales data. After uploading the dataset, he asks \tool to help him build a demand forecasting pipeline. As the pipeline runs, \tool presents each stage using a multi-level semantic representation interface. Johnny adjusts the semantic slider from L1 to L5 (Figure~\ref{fig:UI}\mycircle{A}) and sees that each step card progressively expands from concise summaries to more detailed logs of the same stage. Since he is not an experienced programmer, Johnny stays at L2, which summarizes each step in natural language. This helps him understand the overall progress of the workflow.

As the workflow progresses, Johnny notices that the \textit{Exploratory Data Analysis} step has completed. The L2 summary (Figure~\ref{fig:UI}\mycircle{A}) highlights \textit{strong annual seasonality and 500 store-item combinations, each representing a separate time series}. This gives Johnny a useful overview, but also raises a question. He drags the L2 summary into the input box (Figure~\ref{fig:UI}\mycircle{C}) and types, ``\textit{Is this seasonal pattern consistent across all store-item groups?}'' (Figure~\ref{fig:UI}\mycircle{D}). \tool explains that although the overall dataset shows clear annual seasonality, the pattern varies across store-item combinations, and some volatile series exhibit weaker or noisier seasonal signals. This helps Johnny refine his interpretation: the pattern is uneven across groups.


As the pipeline continues, Johnny notices that \textit{Feature Engineering} has just completed and the agent has moved on to \textit{Baseline Modeling}. A warning badge appears on the completed \textit{Feature Engineering} card (Figure~\ref{fig:UI}\mycircle{B}). He opens the warning panel and clicks ``\textit{Why this warning?}'' (Figure~\ref{fig:warning}\mycircle{1}) to understand what happened. \tool explains that the step used a global \texttt{dropna()} operation, which may remove rows near the beginning of each store-item sales history, even when those rows could still be useful (Figure~\ref{fig:warning}\mycircle{2}). Johnny then clicks the \texttt{feature\_engineering.py} file icon, which takes him to the relevant line in the script. Although he does not fully understand the script, he confirms this line calls \texttt{dropna()}. He selects the line and asks MUSE to explain it. \tool confirms that the operation removes every row with missing values, including early records without enough history for rolling features.

After understanding the issue, Johnny clicks the \textit{Revise} button (Figure~\ref{fig:warning}\mycircle{1}). \tool{} then presents three revision options (Figure~\ref{fig:warning}\mycircle{3}). Johnny chooses \textit{``Keep rows with partial history''} because it best matches his understanding of the warning—some rows may still be useful even without enough history for every feature. By comparison, the other two options feel less consistent with the previous explanation.

After the workflow completes, Johnny opens \texttt{final\_report.md} to inspect the modeling results. He notices that the \textit{Random Forest} model is recommended based on an R\textsuperscript{2} score of 0.9296. Since he is unfamiliar with this metric, he first asks \tool what R\textsuperscript{2} means and why it matters for this recommendation. \tool explains that R\textsuperscript{2} measures how well the model's predictions match the observed sales values. He then highlights the metric in the file and chooses \textit{Help me verify} (Figure~\ref{fig:UI}\mycircle{E}). \tool then presents several verification options (Figure~\ref{fig:verification}\mycircle{1}). Johnny chooses to recompute the metric because this feels like the fastest way to confirm the value is correct. \tool then generates a verification script to recompute the R\textsuperscript{2} score. When the result shows that the R\textsuperscript{2} score was computed correctly (Figure~\ref{fig:verification}\mycircle{2}), Johnny feels assured about the reported value.
}
\begin{table*}[h!]
\centering
\small
\setlength{\tabcolsep}{4pt}

\begin{tabular}{p{0.9cm} p{3.2cm} p{1.1cm} p{3.3cm} p{3.9cm} p{1.8cm}}
\toprule
\textbf{TID} & \textbf{Dataset} & \textbf{Difficulty} & \textbf{Task Goal} & \textbf{Key Features} & \textbf{Target} \\
\midrule

\textbf{T1} & \textbf{Food Delivery Time} & Easy &
Predict delivery time for reliable ETA estimation. &
Distance, weather, traffic, vehicle type, etc. &
Delivery time \\

\textbf{T2} & \textbf{Demand Forecasting} & Easy &
Forecast product demand for inventory planning. &
Date, store, item &
Sales \\

\textbf{T3} & \textbf{Telco Customer Churn} & Medium &
Identify customers likely to cancel subscriptions. &
Tenure, Internet service, contract, monthly charges, etc. &
Churn (Yes/No) \\

\textbf{T4} & \textbf{Customer Lifetime Value} & Medium &
Predict future customer revenue from historical transactions. &
Quantity, unit price, customer ID, invoice date, etc. &
Future revenue \\

\textbf{T5} & \textbf{MovieLens 100k} & Hard &
Predict user preferences for unseen movies. &
User ratings, movie metadata, genre indicators, demographics &
Rating (1--5) \\

\textbf{T6} & \textbf{Online News Popularity} & Hard &
Predict article popularity before publication. &
Content statistics, keyword metrics, topic features, sentiment, etc. &
Shares \\

\bottomrule
\end{tabular}
\caption{\new{User study tasks.}}
\label{tab:dataset_summary}
\vspace{-20pt}
\end{table*}

\section{USER STUDY DESIGN}
\label{sec:evaluation}
We conducted a between-subjects user study with 15 participants from diverse academic backgrounds, including Business Analytics, Computer Science, Finance, Architecture, Aeronautics and Astronautics, Botany and Plant Pathology, Statistics, Education, Materials Science and Engineering, Physics, and Mechanical Engineering. Table~\ref{tab:participant_assignment} reports detailed demographic information.
\begin{table}[h]
\centering
\setlength{\tabcolsep}{5pt}
\resizebox{\columnwidth}{!}{%
\begin{tabular}{c c l c c}
\toprule
\textbf{ID} & \textbf{Condition} & \textbf{Background} & \textbf{Expert} & \textbf{Gender} \\
\midrule
R1 & A & Business Analytics & Yes & Male \\
R2 & A & Computer Science & Yes & Male \\
R3 & A & Finance & No & Male \\
R4 & A & Architecture & No & Male \\
R5 & A & Finance & No & Male \\
L1 & B & Aeronautics \& Astronautics & No & Male \\
L2 & B & Computer Science & Yes & Male \\
L3 & B & Botany \& Plant Pathology & No & Female \\
L4 & B & Statistics & No & Male \\
L5 & B & Computer Science & Yes & Male \\
M1 & C & Education & No & Male \\
M2 & C & Materials Science \& Eng. & Yes & Female \\
M3 & C & Computer Science & Yes & Male \\
M4 & C & Physics & No & Male \\
M5 & C & Mechanical Engineering & No & Male \\
\bottomrule
\end{tabular}%
}
\caption{\new{Participant assignment, background, self-reported agent expertise, and gender.}}
\label{tab:participant_assignment}
\vspace{-15pt}
\end{table}

We compared user performance across three conditions with different levels of abstraction and interaction support.

\noindent\textbf{Condition A (Raw Logs).} In this condition, users see the underlying agent's raw execution outputs, including code edits, reasoning traces, intermediate plans, tool calls, and runtime messages. This design resembles existing agent execution interfaces~\cite{nam2025mle, zhang2025deepanalyze} that expose low-level traces directly. Participants can inspect these logs and provide feedback through a chat interface.

\noindent\textbf{Condition B (DiLLS).} This condition represents a stronger baseline inspired by prior work~\cite{sheng2026dills}. The interface explains the agent's behavior through three-layered summaries that expand from high-level \textit{activities} to lower-level \textit{actions} and \textit{operations}. Similar to Condition C, Condition B surfaces warnings for inspection using the same warning detection method. However, these warnings do not support in-situ action on the workflow. Users must manually recover the relevant context and switch to external editors to inspect, verify, or modify the workflow.

\noindent\textbf{Condition C (\tool).} \tool\ converts agent behavior traces into a multi-level semantic representation of the underlying DS workflow. The interface is structured around DS workflow semantics rather than individual agent behaviors, allowing users to move from high-level workflow overviews to step-specific details. \tool\ proactively monitors the workflow, surfaces localized warnings for suspicious steps, supports contextualized question answering, and enables in-context intervention by allowing users to reference workflow steps to verify agent's behavior, inspect suspicious outputs, and revise the workflow at their preferred level of abstraction.
\new{

We selected six representative DS tasks from Kaggle~\cite{kaggle} for the user study. We chose Kaggle tasks since Kaggle is a primary benchmark source for SOTA data science agents~\cite{nam2025mle, zhang2025deepanalyze}, and WaitGPT~\cite{xie2024waitgpt} also adopts Kaggle tasks in its user study. We are aware that some classic Kaggle tasks, such as the Titanic survival prediction task, are unrealistic and may have been memorized by LLMs. Therefore, we selected tasks that resemble real-world scenarios (e.g., food delivery time prediction, customer churn prediction) rather than classic benchmark tasks. The six tasks are detailed in Table~\ref{tab:dataset_summary}.

To expose participants to agent behaviors with varying error characteristics, the study included both a weaker model (Claude Haiku 4.5) and a stronger model (Claude Opus 4.6). The assignment of conditions, tasks, and models was counterbalanced across participants. During the study, we did not tell the participants which model they were using. Furthermore, our error analysis of agent behavior on these tasks shows that agents make data science mistakes resembling those of human practitioners, as discussed in Section~\ref{warningtypedist}.

At the beginning of each session, participants received a brief introduction and tutorial. They then completed two tasks under their assigned condition, with 30 minutes allotted per task. For each task, participants received a description of the dataset and the prediction task, and were asked to craft their own prompts rather than copying the provided descriptions. A task was considered successful if the participant produced a runnable DS pipeline within the time limit, and the experimenters manually verified that the pipeline performed appropriate preprocessing and feature engineering, used a suitable model, and trained and evaluated it correctly. Any mistakes in these steps were counted as a task failure. We did not require the agent to produce the best-performing model, since model optimization is too time-consuming for a 30-minute study and would make the evaluation overly rigid.} After completing both tasks, participants filled out a post-task questionnaire and took part in a brief semi-structured interview. Each session lasted 72 minutes on average, and participants received a \$50 Amazon gift card as compensation. Further details on the study procedure, measures, data analysis, and interview protocol are provided in Appendix~\ref{sec:procedure}, Appendix~\ref{sec:measures}, Appendix~\ref{sec:data_analysis}, and Appendix~\ref{sec:interviewquestions}.


\definecolor{secbg}{HTML}{F7F6F3}
\definecolor{success}{HTML}{2E7D32}
\definecolor{failure}{HTML}{C62828}

\begin{table*}[t]
\centering
\footnotesize
\setlength{\tabcolsep}{3pt}
\renewcommand{\arraystretch}{1}

\begin{tabular}{@{} l *{6}{c} *{6}{c} *{6}{c} @{}}

\toprule

& \multicolumn{6}{c}{\textbf{Raw Logs}}
& \multicolumn{6}{c}{\textbf{DiLLS}}
& \multicolumn{6}{c}{\textbf{MUSE}} \\

\cmidrule(lr){2-7} \cmidrule(lr){8-13} \cmidrule(lr){14-19}

& R1 & R2 & R3 & R4 & R5 & \textbf{Avg}
& L1 & L2 & L3 & L4 & L5 & \textbf{Avg}
& M1 & M2 & M3 & M4 & M5 & \textbf{Avg} \\

\midrule

\rowcolor{secbg}
\multicolumn{19}{@{}l}{\textbf{Task 1}} \\

\textit{Task ID}
& T1 & T1 & T2 & T3 & T3 &
& T1 & T1 & T2 & T4 & T5 &
& T1 & T2 & T2 & T3 & T4 & \\

\textit{Model}
& Opus & Haiku & Opus & Opus & Opus &
& Opus & Opus & Opus & Opus & Opus &
& Haiku & Opus & Haiku & Opus & Opus & \\

\textit{Time (min)}
& 26 & 21 & 30 & 30 & 30 & \textbf{27.4}
& 21 & 10 & 30 & 16 & 26 & \textbf{20.6}
& 14 & 17 & 16 & 25 & 30 & \textbf{20.4} \\

\textit{Attempts}
& 1 & 1 & 1 & 1 & 1 & \textbf{1.0}
& 1 & 1 & 1 & 2 & 1 & \textbf{1.2}
& 1 & 1 & 1 & 1 & 2 & \textbf{1.2} \\

\textit{Result}
& \textcolor{success}{S} & \textcolor{success}{S} & \textcolor{failure}{F} & \textcolor{failure}{F} & \textcolor{failure}{F} &
& \textcolor{success}{S} & \textcolor{success}{S} & \textcolor{failure}{F} & \textcolor{success}{S} & \textcolor{success}{S} &
& \textcolor{success}{S} & \textcolor{success}{S} & \textcolor{success}{S} & \textcolor{success}{S} & \textcolor{failure}{F} & \\

\addlinespace[4pt]

\rowcolor{secbg}
\multicolumn{19}{@{}l}{\textbf{Task 2}} \\

\textit{Task ID}
& T4 & T6 & T5 & T4 & T6 &
& T2 & T3 & T4 & T5 & T6 &
& T5 & T3 & T6 & T5 & T6 & \\

\textit{Model}
& Haiku & Opus & Haiku & Haiku & Haiku &
& Haiku & Haiku & Haiku & Haiku & Haiku &
& Opus & Haiku & Opus & Haiku & Haiku & \\

\textit{Time (min)}
& 22 & 23 & 20 & 30 & 30 & \textbf{25.0}
& 25 & 10 & 14 & 12 & 22 & \textbf{16.6}
& 9 & 9 & 16 & 22 & 12 & \textbf{13.6} \\

\textit{Attempts}
& 1 & 1 & 1 & 1 & 1 & \textbf{1.0}
& 3 & 1 & 1 & 1 & 1 & \textbf{1.4}
& 1 & 1 & 2 & 1 & 1 & \textbf{1.2} \\

\textit{Result}
& \textcolor{success}{S} & \textcolor{success}{S} & \textcolor{success}{S} & \textcolor{failure}{F} & \textcolor{failure}{F} &
& \textcolor{success}{S} & \textcolor{success}{S} & \textcolor{success}{S} & \textcolor{success}{S} & \textcolor{success}{S} &
& \textcolor{success}{S} & \textcolor{success}{S} & \textcolor{success}{S} & \textcolor{success}{S} & \textcolor{success}{S} & \\

\bottomrule

\end{tabular}

\caption{Per-participant assignments and  outcomes by condition.
\textnormal{\textcolor{success}{S}}~=~success,
\textnormal{\textcolor{failure}{F}}~=~failure.}
\label{tab:condition_outcomes}
\vspace{-20pt}
\end{table*}

\section{USER STUDY RESULTS}
\label{sec:results}

\subsection{User Performance}


\subsubsection{{\bf \textit{Overall Performance}}}
As shown in Table~\ref{tab:condition_outcomes}, participants achieved a 90\% success rate in both Conditions B and C, compared with only 50\% in Condition A. Furthermore, participants in Condition C had the shortest average completion time per task (17 min), representing a 35\% reduction compared to Condition A and a 9\% reduction compared to Condition B. \new{An ANOVA test showed that the difference in completion time across the three conditions was statistically significant ($F(2,12)=4.71$, $p=.031$, $\eta^2=.44$). Tukey HSD post-hoc analysis indicated that \tool\ significantly reduced completion time relative to Condition A ($p=.035$), while the reduction relative to Condition B was not statistically significant ($p=.875$).} Participants in Condition C made an average of 1.20 attempts per task, compared to 1.00 in Condition A and 1.30 in Condition B, where an \textit{attempt} refers to restarting a task from scratch. \new{The difference in attempts was not statistically significant ($F(2,12)=1.27$, $p=.315$, $\eta^2=.18$).} These results suggest that \tool\ helped participants complete tasks more efficiently while maintaining a high success rate.


\subsubsection{{\bf \textit{User Confidence}}}
\begin{figure}[ht!]
    \centering
    \includegraphics[
        width=\columnwidth
    ]{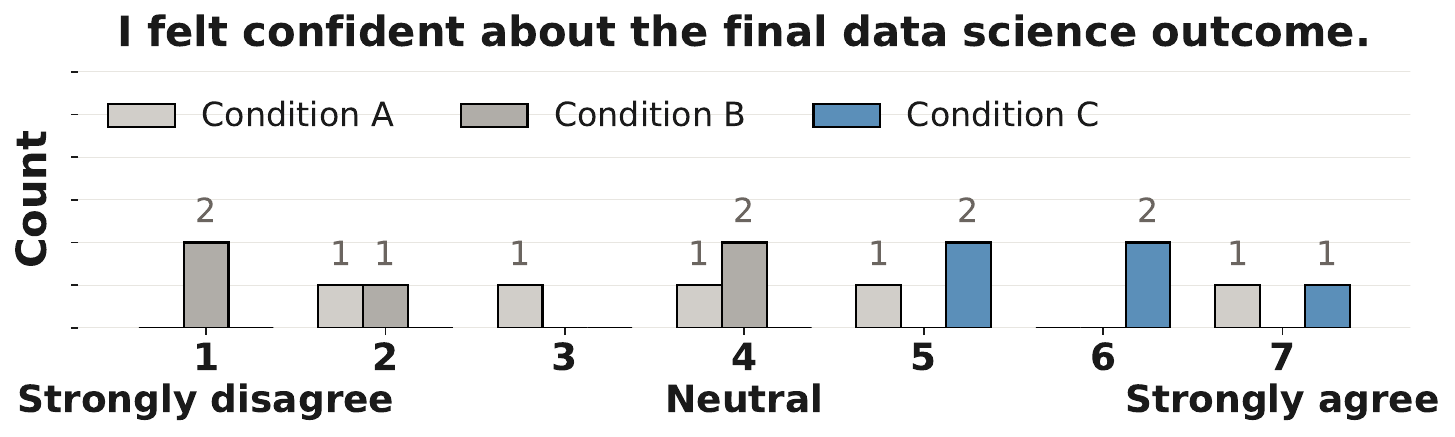}
    \caption{ User confidence in the final data science outcomes under conditions A, B, and C.
    }
    \label{fig:userconfidence}
\end{figure}

As shown in Figure~\ref{fig:userconfidence}, participants reported the highest confidence  in Condition C (\textit{Mean}: 4.20 vs.~2.60 vs.~5.80). An ANOVA test showed that the differences in users' confidence across the three conditions were statistically significant ($p$ = .0265). This result suggests that {\tool} better supported users in building confidence in the DS outcomes. To better understand why participants felt more confident when using {\tool}, we further analyzed their perceptions in Section~\ref{sec:userPerceptions}.


\begin{figure}[ht!]
    \centering
    \includegraphics[
        width=\columnwidth
    ]{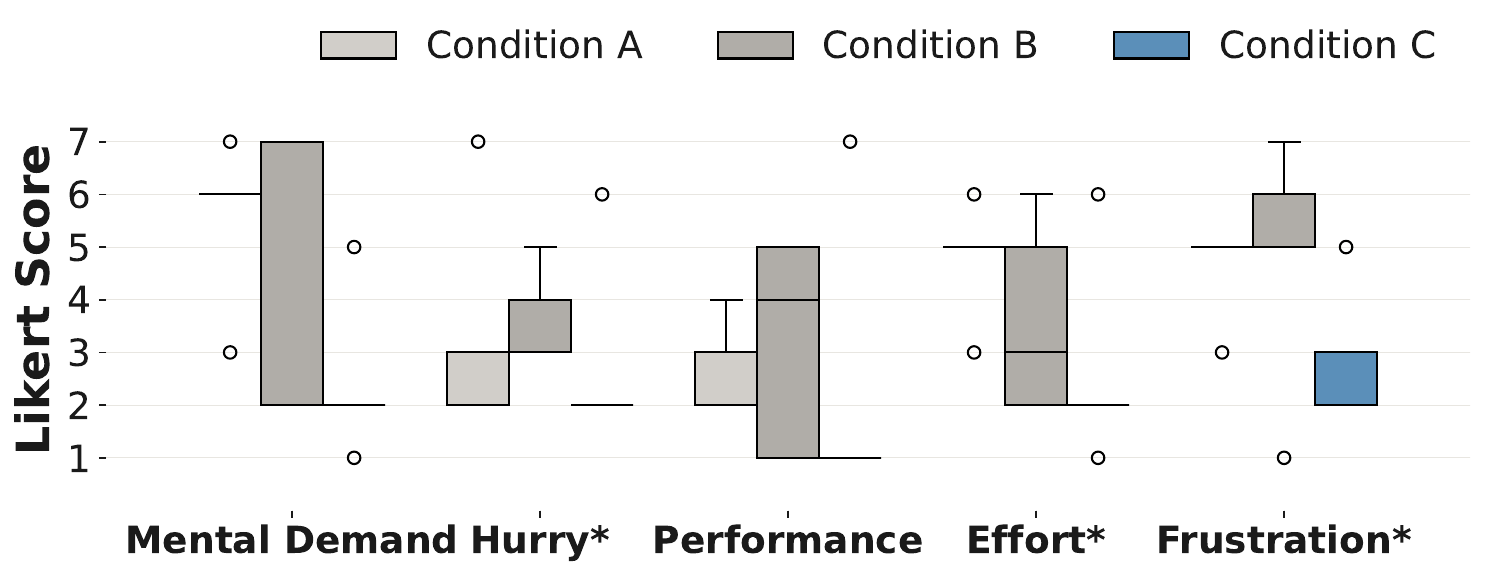}
    \caption{
    User responses to the NASA TLX questionnaire (*: $p < .05$ based on the ANOVA test).
    }
    \label{fig:nasatlx}
\end{figure}
\begin{figure*}[ht!]
    \centering

    \includegraphics[
        width=0.9\textwidth,
        trim=0.7cm 0.1cm 0.7cm 0.3cm,
        clip
    ]{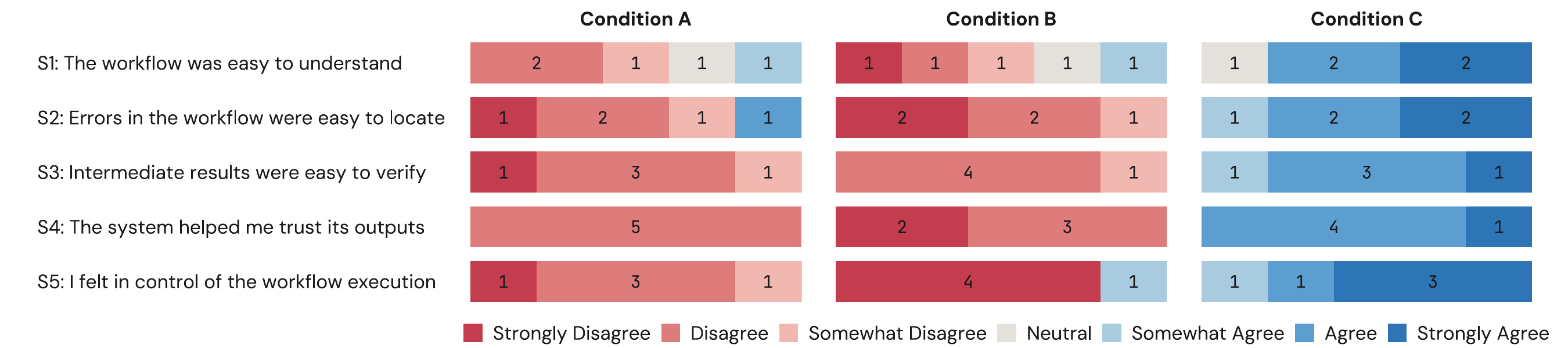}
    \vspace{-10pt}
    \caption{7-point Likert scale evaluations (1 = strongly disagree, 7 = strongly agree) of five comparison statements: S1 (understanding), S2 (error localization),  S3 (verification), S4 (trust), and S5 (control).
    }
    \label{fig:subjectiveMeasure}
    \vspace{-10pt}
\end{figure*}
\subsubsection{{\bf \textit{Cognitive Overhead}}}
As shown in Figure~\ref{fig:nasatlx}, we found significant differences across conditions for \textit{Hurried}, \textit{Effort}, and \textit{Frustration} (ANOVA: $p = .0165$, $.0298$, and $.0428$, respectively). In contrast, we did not detect significant differences for \textit{Demand} or \textit{Performance} ($p = .1516$ and $.1196$, respectively).

\subsubsection{{\bf \textit{Error Handling}}}
Condition A did not surface explicit warnings about potential agent mistakes. We compare outcomes only between the two assisted conditions. While Condition B and Condition C use the same warning detection method, the total number of warnings detected in each condition is slightly different due to the variations in task assignment and the inherent randomness in LLM inference. In total, 39 warnings were detected in Condition B and 36 warnings were detected in Condition C. We found that participants resolved substantially more warnings in Condition C than in Condition B, fixing an average of 2.8 warnings compared to 0.8. This difference was statistically significant according to an ANOVA test ($p$ = .0118). This suggests that \tool's in-situ repair features made it easier for participants to resolve surfaced warnings.

\new{
\subsubsection{{\bf \textit{Warning Type Distribution}}}\label{warningtypedist}

We found that agents made recurring data science mistakes across all tasks, with the most frequent categories being \textit{Unverified Schema Assumptions} (56\%), \textit{Missing Data Not Inspected} (16\%), and \textit{Single-Split Evaluation} (12\%). We also examined warning frequency by model assignment: Haiku triggered 27\% more warnings than Opus, suggesting that weaker models may be more prone to exhibiting suspicious behaviors. In addition, participants were not told which model they were using, and we did not find evidence that they could reliably tell whether they were interacting with a weaker or stronger model.
}

\new{
\subsubsection{{\bf \textit{User Behavior by Warning Type}}}

We further examined how participants reacted to different types of warnings. Methodological validation warnings, such as \textit{single-split evaluation}, were most consistently acted upon, with 7 revision requests. Domain and schema interpretation warnings attracted less user attention (2 revised, 1 queried). This suggests that users were more inclined to act on warnings with an actionable fix, while warnings requiring deeper domain judgment were more often left unresolved.
}

\new{
\subsubsection{{\bf \textit{From Linear Search to Localized Inspection}}}
Our analysis suggests that MUSE reshaped monitoring from an act of \textit{linear searching} into an act of \textit{localized inspection}. Under the raw logs condition, users relied on linear scrolling to locate relevant information (147 scrolls/task); under MUSE, scrolling dropped substantially (61 scrolls/task). The semantic-level slider was the most frequent interaction in Condition C, with 162 adjustments distributed across every workflow stage. We found that these adjustments were often triggered by warnings, newly completed steps, or checks before invoking \textit{Revise}. Together, these findings suggest that once execution is organized into semantic chunks with overview-to-detail navigation, users stop scanning raw logs and instead navigate directly to the semantic level and artifact most relevant to their current goal.
}

\subsection{User Perceptions of Different Conditions}
We analyzed participants' perceptions across five comparison statements (\textbf{S1--S5}) shown in Figure~\ref{fig:subjectiveMeasure}. In the quotes below, participants are identified as \textbf{R1--R5}, \textbf{L1--L5}, and \textbf{M1--M5}, corresponding to Condition A, Condition B, and Condition C, respectively.
\label{sec:userPerceptions}

\subsubsection{{\bf \textit{Multi-level Representations Helped Users Better Understand Agent Workflows}}} 
Participants felt that agent workflows were easier to understand in Condition C than the other two conditions \new{(S1, $F(2,12)=7.40$, $p=.008$, $\eta^2=.55$; Tukey HSD: \tool\ vs.\ A/B, $p=.019/.013$).} Several participants noted that the multi-level semantic representation helped them follow workflow progress (M1, M2, M3, and M4). M2 said, ``\textit{I preferred the hierarchical, step-by-step structure because it made the workflow easier to follow.}'' By contrast, participants in Condition A frequently described the raw logs as dense and difficult to interpret. R4 noted that ``\textit{there is indeed a lot of information, and it is hard for you to understand in a short time what exactly it is doing.''} Condition B provided some support for understanding the workflow by presenting layered summaries of individual agent behavior, but many users still found these explanations too developer-oriented. L3 noted that ``\textit{the high-level action summaries were helpful, but the lower-level operations were still too cryptic for me to understand the overall workflow, since I do not have experience developing agents.}''
\begin{figure*}[ht!]
    \centering

    \includegraphics[
        width=\textwidth,
        trim=0.4cm 0.2cm 0.2cm 0.7cm,
        clip
    ]{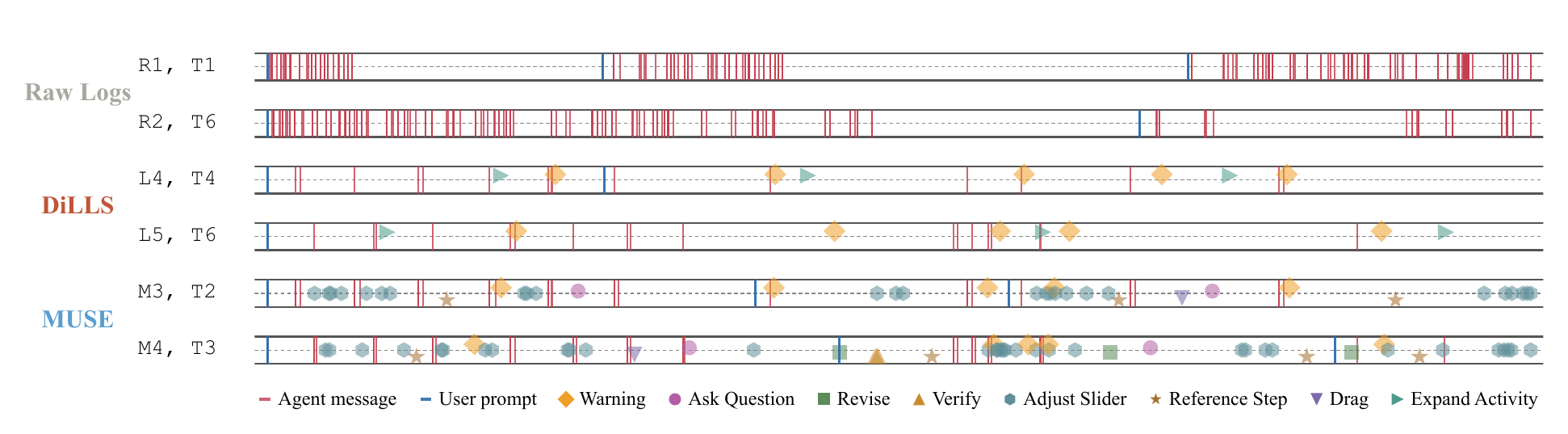}
    \vspace{-20pt}
    \caption{ Representative user event timelines across conditions in normalized time.
    }
    \label{fig:eventsDistribution}
    \vspace{-15pt}
\end{figure*}
\subsubsection{{\bf \textit{Localized Warnings Helped Users Identify Suspicious Behavior}}} 
Participants in Condition C felt that it was easier to identify suspicious steps than the other two conditions \new{(S2, $F(2,12)=15.65$, $p<.001$, $\eta^2=.72$; Tukey HSD: \tool\ vs.\ A/B, $p=.004/<.001$)}. Several participants highlighted that the localized warnings helped them quickly focus on suspicious parts without needing to inspect the entire execution trace (M1, M2, M3, and M5). They described the warnings as ``\textit{obvious,}'' ``\textit{localized,}'' and ``\textit{actionable,}'' which supported ``\textit{quick troubleshooting.}'' By contrast, participants in Condition A often struggled to detect issues. R3 noted that \textit{``some intermediate steps flashed by too quickly to catch any errors, you have to wait for it to stop then read everything from scratch.''} Condition B surfaced warnings, but these warnings lacked in-situ support for follow-up repair. Consequently, users often needed to context-switch to external editors to address the underlying issue (L1, L3, L4, and L5). L5 said, \textit{``Although the errors showed some hints, I still need to go to the editor to identify the source of the errors.''}
 
\subsubsection{{\bf \textit{Verification Scaffolds Helped Users Build Trust in Results}}}
In Condition C, participants found verification easier and trusted the results more than in the other two conditions \new{(S3, $F(2,12)=63.50$, $p<.001$, $\eta^2=.91$; S4, $F(2,12)=194.80$, $p<.001$, $\eta^2=.97$; Tukey HSD: all pairwise comparisons $p<.001$ for both S3 and S4)}. In Condition C, users mentioned that the verification scaffolds supported systematic checking without requiring them to manually reconstruct context. M2 noted, ``\textit{It enhanced my trust in the results,}'' while M3 emphasized that ``\textit{manual verification would waste time, and the scaffolded verification features with actionable options helped me directly verify the intermediate results easily.}'' Conversely, participants in Conditions A and B had only limited verification support, forcing users to manually inspect intermediate outputs. R5 said that ``\textit{the verification process was too complex for beginners.}'' L3 pointed out that relying solely on the UI provided no confidence, stating that users ``\textit{have to dig into backend files or read the raw code to confidently verify any result.}'' Similarly, L1 felt they had ``\textit{no way to verify the results directly.}''

\subsubsection{{\bf \textit{Contextualized Repair Helped Users Steer the Workflow}}} 
Participants in Condition C reported that it was easier to control the agent than in the baseline conditions \new{(S5, $F(2,12)=22.53$, $p<.001$, $\eta^2=.79$; Tukey HSD: all pairwise comparisons $p<.001$)}. In Conditions A and B, participants lacked in-situ control and had to wait for the entire pipeline to finish before writing prompts to steer the agent. L4 noted, \textit{``It seems impossible to modify it midway, because you have to wait for it to finish completely before adding new commands.''} Conversely, \tool\ enabled users to steer the agent by combining contextualized referencing with option-driven repairs. Users could directly drag workflow steps into the chat to specify where to steer, an interaction that M1 found \textit{``much clearer''} than pure text conversations. To support repair, \tool\ automatically suggested actionable options. M2 praised this \textit{``very convenient''} design. M1 and M4 also noted that buttons such as \textit{ ``Revise''} and \textit{``Verify''} reduced the need for prolonged back-and-forth prompting.

\subsection{User Interaction Patterns}

\subsubsection{{\bf \textit{Interaction Patterns Across All Conditions}}} 
 Figure~\ref{fig:eventsDistribution} shows the temporal distribution of events across the three conditions in normalized time\footnote{To preserve readability, we visualize two representative participants from each condition whose timelines reflect common interaction patterns. The complete event timelines for the remaining participants are provided in the supplementary materials.}. In Condition A, participants were exposed to a dense stream of raw logs throughout the session. Accordingly, the interaction traces are dominated by frequent agent messages, reflecting a passive monitoring style with limited opportunities for in-situ intervention. In Conditions B and C, execution traces were instead converted into layered summaries and multi-level representations, enabling users to inspect workflow progress at different levels of abstraction. However, participants in Condition B appeared to spend more effort verifying agent behavior, as they more often needed to open external editors to inspect intermediate results. By contrast, participants using \tool\ engaged in more in-situ interactions during task completion, with 7 Ask actions, 5 Verify actions, and 10 Revise actions observed across the MUSE logs. This suggests a more mixed-initiative workflow that supported iterative checking and repair with less context switching.


\subsubsection{{\bf \textit{Delegators, Collaborators, and Auditors}}} 
Prior work~\cite{gu2024analysts, chen2025code, barke2023grounded, vaithilingam2022expectation} suggests that users exhibit different behaviors when inspecting and verifying AI-generated outputs. We also observed three behavior patterns in our study: \textbf{AI Delegators} (R2, L2, and M5), who largely trusted the agent’s outputs; \textbf{AI Collaborators} (R1, R3, R4, R5, L1, L4, M1, and M2), who selectively scanned agent messages and outputs; and \textbf{AI Auditors} (L3, L5, M3, and M4), who carefully examined intermediate results and system messages. For example, R2, who was an AI delegator, expressed strong trust in the agent, saying, \textit{``I am very confident that AI can do everything right.''} By contrast, R5, who was an AI collaborator, described a more selective strategy: \textit{``I would first skim what it did, and only check more carefully if something looked off.''} L5, who was an AI auditor, described a more exhaustive approach: \textit{``AI will hallucinate, so I would read every detail to make sure it does not mess up.''} Together, these observations suggest that users may need different levels of support for verification in order to build trust in the agent.

\subsubsection{{\bf \textit{The Impact of User Expertise}}} 

We categorized participants into two groups based on technical background: experts, who had more than five years of computer science training or agent-development experience ($N=6$), and novices, who did not ($N=9$). We compared task performance between the two groups in terms of completion time and number of attempts. On average, experts completed tasks in 18 minutes and 11 seconds, whereas novices completed them in 22 minutes and 11 seconds, although this difference was not statistically significant (ANOVA test: $p$ = .241). The average number of attempts was also similar: experts made 1.08 attempts on average, compared with 1.22 for novices, and this difference was likewise not statistically significant (ANOVA test: $p$ = .413). Overall, these results suggest that prior programming or agent-development experience did not significantly affect users' ability to complete tasks with AI agent tools.


\section{Discussion}

\subsection{Design Implications}

\subsubsection{{\bf \textit{The Shifting Error Landscape in Agentic Systems}}}
We found that in data science tasks, current agentic systems rarely made simple mistakes such as syntax errors or system crashes, as reported in prior work~\cite{gu2024analysts, zhang2025which, cemri2025multi}, even when using a weaker model (Claude Haiku 4.5). However, users were often presented with results that appeared complete but still contained many \textit{silent errors}, including questionable assumptions, inappropriate data processing, or logically flawed intermediate steps that did not trigger execution failures or crashes. One promising direction for future work is to enable agents to ask users before making high-impact decisions, such as removing records or interpreting under-specified intent. However, simply asking for clarification may not be enough, as this could shift responsibility back to end-users with limited knowledge and create an \textit{information barrier}~\cite{ko2004six}. Future systems may therefore need to accompany such questions with contextual information, such as why the action was proposed, what data will be affected, and how the choice may influence downstream results.

\subsubsection{{\bf \textit{Adaptive Abstraction for Agentic Systems}}}
In previous LLM-based DS agents~\cite{zhang2025deepanalyze, nam2025mle}, outputs are often verbose and unstructured. In this work, we take an initial step toward addressing this challenge by introducing a multi-level semantic abstraction hierarchy to represent system behavior. However, no single abstraction design is likely to fit all users. Future work could explore more flexible mechanisms that allow users to customize how abstractions are structured. One alternative is to enable \textit{malleable user interfaces}~\cite{min2025malleable}, which allow users to reshape how information is organized and represented. Another direction is to incorporate \textit{just-in-time objectives}~\cite{lam2026jitObjectives}, where the system infers a user’s immediate goal from interaction context and dynamically generates specialized views or tools to support that goal. However, these adaptive abstractions must be designed carefully, as dynamically changing the content or representation of information may reduce interface consistency, make system behavior harder to predict, and hinder users’ ability to form stable mental models of the system~\cite{nielsen1994enhancing}.

\subsection{Limitations}
Our user study was conducted in a controlled setting with six DS tasks, which may not fully capture the complexity of real-world DS practice. In addition, while our participants represented diverse academic backgrounds, many were not professional developers. Therefore, our findings primarily reflect how MUSE supports less-experienced users, rather than its effectiveness for expert practitioners. Moreover, MUSE's monitoring layer currently relies on pre-defined warning heuristics and therefore may not capture the broader spectrum of errors that arise in practice.

\section{Conclusion}
We presented \tool{}, an interactive meta-agent designed to help users understand and steer LLM-powered data science agents. By dynamically restructuring low-level execution traces into multi-level semantic representations and supporting contextualized questioning, verification, and in-situ revision, \tool provides users with more accessible ways to inspect and intervene in agent-generated workflows. Results from our between-subjects study suggest that \tool can improve task efficiency, increase user confidence, and strengthen users’ trust in the agentic data science workflow.


\bibliographystyle{ACM-Reference-Format}
\bibliography{citations}

\appendix
\onecolumn

\section{Formative Study}

\subsection{Formative Study Procedure}
\label{appendix:formativeProcedure}
We conducted task-based study sessions with each participant. In each session, participants completed two data science tasks using two LLM-powered systems while thinking aloud, followed by a semi-structured interview. Each session lasted approximately 60 minutes.

At the beginning of the session, we introduced the study setting and provided a brief overview of the two systems used in the study: MLE-STAR~\cite{nam2025mle}, an agentic system for machine learning workflows, and Cursor~\cite{cursor2024}, a general-purpose AI coding assistant. Participants then worked on two representative Kaggle-based data science tasks: a tabular prediction task using the California Housing Prices dataset~\cite{california_housing_kaggle} and a prediction task based on Optiver Realized Volatility Prediction~\cite{optiver_kaggle}. To reduce order effects and tool-specific bias, system--task pairings were counterbalanced across participants. We selected these systems to capture different forms of LLM support for data science work, ranging from a more autonomous workflow agent to a general-purpose coding assistant.

During task execution, participants were encouraged to think aloud by verbalizing their reasoning, expectations, and any difficulties they encountered while interacting with the systems. After completing the tasks, we conducted follow-up semi-structured interviews to further probe their experiences. The interview questions focused on:
\begin{itemize}
    \item How they worked with the systems during the tasks.
    \item What aspects of the interaction they found helpful or frustrating.
    \item How they interpreted intermediate outputs.
    \item What they wished they could better understand or control during the process.
\end{itemize}

All sessions were screen-recorded and transcribed for analysis. We conducted an inductive thematic analysis following standard qualitative methods. Two researchers independently coded the transcripts and interaction traces, iteratively discussed discrepancies, and developed a shared codebook. Through this process, we identified recurring patterns and synthesized them into higher-level themes, which informed the user needs and design rationale of \tool.

\section{User Study}

\subsection{User Study Tasks}
\label{sec:tasks}

We selected six representative data science tasks from Kaggle~\cite{kaggle} for the user study, covering a range of predictive modeling scenarios, including regression, classification, time-series forecasting, and recommendation. As summarized in Table~\ref{tab:dataset_summary}, the tasks vary in domain, modeling objective, and data characteristics, allowing us to evaluate \tool under diverse workflow contexts. To reflect different levels of workflow complexity, we categorized the six tasks into three difficulty levels: easy (T1--T2), medium (T3--T4), and hard (T5--T6). This categorization was determined based on two practical factors that affect agent execution difficulty: (1) the number and complexity of input features, and (2) the dataset size, which influences runtime and the amount of processing required. In general, tasks with fewer features and smaller datasets tend to require simpler preprocessing and shorter execution time, whereas tasks with more features or larger files often involve more extensive data handling and longer runtime.

\subsection{User Study Procedure}
\label{sec:procedure}

Each participant took part in a scheduled 80-minute individual study session. At the beginning of the session, participants were introduced to the study setting and received a brief tutorial on the interface. To reduce demand characteristics, we informed participants that the study compared different interfaces for interacting with an AI-powered data science system, without revealing which condition corresponded to our proposed system.

Participants then completed two tasks under their assigned condition. The order of tasks was counterbalanced across participants. For each task, participants received a description of the dataset and the prediction task, and were asked to craft their own prompts rather than copying the provided descriptions. Participants were asked to think aloud while interacting with the system. A task was considered successful if the participant produced a runnable data science pipeline within 30 minutes, and the experimenters manually assessed the final agent trajectory, examining whether the agent performed appropriate preprocessing and feature engineering, selected a suitable model, and trained and evaluated it correctly; any mistakes in these steps were considered a task failure. We did not require the agent to produce the best-performing model, since model optimization is too time-consuming for a 30-minute study and would make the evaluation overly rigid.

After completing both tasks, participants filled out a post-task survey. Finally, they participated in a brief semi-structured interview. Participants received a \$50 Amazon gift card as compensation for their time.

\subsection{User Study Measures}
\label{sec:measures}

We collected both quantitative and qualitative data.

\noindent\textbf{Performance Measures.} We measured task completion time, task success, and the number of task restarts. A task was considered successful if the participant produced a runnable data science pipeline that generated acceptable predictions for the target variable within 30 minutes. We also recorded whether each task was completed within the time limit.

\noindent\textbf{Subjective Measures.} After each task, participants rated their confidence in the final result on a 7-point Likert scale. To assess perceived workload, we administered the NASA TLX questionnaire~\cite{hart1988development}. In addition, participants rated five comparison statements on a 7-point Likert scale: S1 (the workflow was easy to understand), S2 (errors in the workflow were easy to locate), S3 (intermediate results were easy to verify), S4 (the system helped me trust its outputs), and S5 (I felt in control of the workflow execution).

\noindent\textbf{Behavioral Measures.} We logged participants' interactions with the interface to analyze how they used different features under each condition. These interactions included actions such as adjusting semantic levels, inspecting warnings, asking contextualized questions, dragging or referencing workflow steps in the chat, revising flagged steps, invoking verification, and answering system follow-up prompts.

\noindent\textbf{Qualitative Data.} We collected think-aloud data during task execution, open-ended post-task responses, and post-study interview feedback to understand how participants interpreted system behavior, diagnosed problems, verified results, and perceived control under different conditions.

\subsection{User Study Data Analysis}
\label{sec:data_analysis}
For quantitative measures, we used ANOVA to compare performance and subjective ratings across conditions. We report descriptive statistics and $p$-values across all analyses. For qualitative data, we used inductive thematic analysis. Two researchers independently reviewed the think-aloud data, post-task responses, interview transcripts, and interaction traces, iteratively developed a shared codebook, discussed disagreements, and refined the final themes through consensus.

\subsection{User Study Interview Questions}
\label{sec:interviewquestions}

After each condition, we conducted a semi-structured interview. We asked participants five main questions:

\begin{itemize}
    \item \textbf{Understanding:} Did you feel you understood what the system was doing?
    \item \textbf{Diagnosis:} When the result seemed suspicious, how did you diagnose the problem?
    \item \textbf{Verification:} How did you verify whether the final result was correct?
    \item \textbf{Revision:} After finding a problem, how did you tell the system what to revise?
    \item \textbf{Control:} Did you feel able to control the system’s behavior during use?
\end{itemize}

For each question, we asked follow-up probes about concrete difficulties, such as not knowing where to look, what information to inspect first, wanting to verify but not knowing how, difficulty specifying revisions without additional context, or wanting to intervene mid-process but not knowing how.

\clearpage
\section{Underlying System Architecture}
\label{appendix:underlyingDSAgent}

\begin{figure}[ht]
    \centering
    \includegraphics[width=0.7\linewidth]{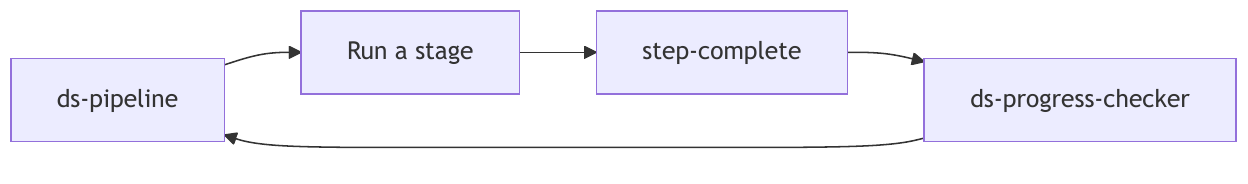}

    \caption{High-level control loop of the skill-based data science pipeline. The orchestrator executes or dispatches a stage, emits a completion signal, and then invokes the progress checker to determine the next action.}

    \label{fig:ds-pipeline}
    
\end{figure}

The underlying data science pipeline is implemented by instantiating a Claude Code agent through the Claude SDK~\cite{ClaudeSDK} and equipping that agent with a set of custom skills~\cite{anthropic_claude_code_skills}. In the current prototype, control is organized around an execution loop consisting of \texttt{ds-pipeline}, \texttt{step-complete}, and \texttt{ds-progress-checker}, while stage-specific analytical work is delegated to \texttt{ds-cleaning}, \texttt{ds-eda}, \texttt{ds-feature-engineering}, and \texttt{ds-modeling}. As shown in Figure~\ref{fig:ds-pipeline}, \texttt{ds-pipeline} executes or dispatches a stage, emits a completion signal, and then invokes the progress checker to determine the next action.

\subsection{Relationship Between \tool and \texttt{ds-pipeline}}

\begin{figure}[ht]
    \centering
    \includegraphics[width=0.9\linewidth]{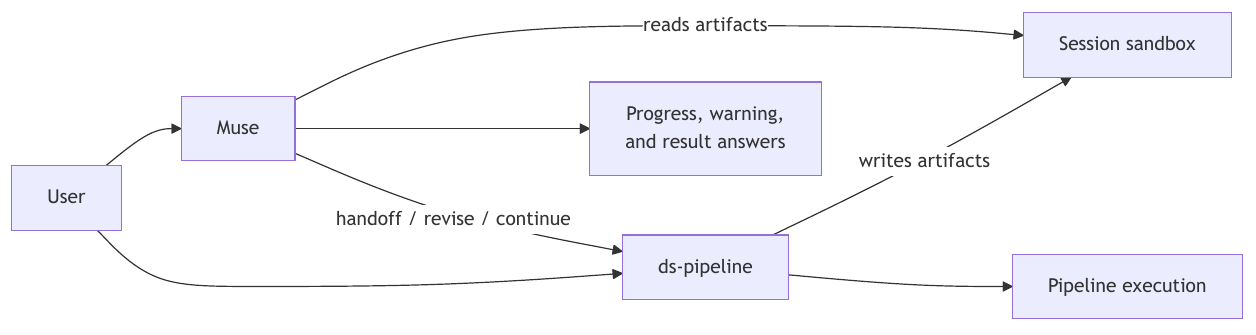}

\caption{Relationship between \tool and \texttt{ds-pipeline}. \texttt{ds-pipeline} advances the staged workflow and writes artifacts to the session sandbox, while \tool reads those artifacts to answer user questions about progress, warnings, and results. When the interaction requires revising or continuing execution, \tool hands control back to \texttt{ds-pipeline}.}

    \label{fig:muse_ds_pipeline_relationship}
    
\end{figure}

The prototype also includes \tool, a dedicated read-only agent implemented using a separate Claude agent runtime. \tool operates alongside \texttt{ds-pipeline} rather than replacing it. While \texttt{ds-pipeline} is responsible for advancing the staged workflow and writing artifacts to the session sandbox, \tool reads those artifacts to answer user questions about progress, warnings, and verification results. When the interaction requires revising or continuing execution, \tool hands control back to \texttt{ds-pipeline}. Figure~\ref{fig:muse_ds_pipeline_relationship} illustrates this relationship.

\subsection{Skill: \texttt{ds-pipeline}}
\label{app:skill-pipeline}

\noindent\textbf{Purpose.} An orchestrator for a staged data science pipeline. Sequences seven fixed stages---objective profiling, data-issue detection, cleaning, EDA, feature engineering, baseline modeling, and final reporting---while using \texttt{ds-progress-checker} to control stage transitions and resume logic.

\noindent\textbf{Stages.} (1)~Objective \& Dataset Understanding,
(2)~Detect Data Issues,
(3)~Data Cleaning,
(4)~Exploratory Data Analysis,
(5)~Feature Engineering,
(6)~Baseline Modeling,
(7)~Final Report.

\noindent\textbf{Outputs.} \texttt{objective\_dataset\_report.md}, \texttt{data\_issue\_report.md}, \texttt{data\_cleaned.csv}, \texttt{data\_cleaning\_report.md}, \texttt{eda\_report.md}, EDA figures, \texttt{data\_engineered.csv}, \texttt{feature\_engineering\_report.md}, \texttt{baseline\_model\_report.md}, \texttt{baseline\_metrics.json}, and \texttt{final\_report.md}.

\noindent\textbf{Key Rules.} Distinguish fresh runs from resumes using sandbox artifacts; stages~1 and~7 run inline while stages~2--6 dispatch to sub-skills; invoke \texttt{ds-progress-checker} after Stage~1 and after every subsequent stage; never write artifacts outside \texttt{./sandbox/\$\{CLAUDE\_SESSION\_ID\}/}; allow at most one retry per failed stage and otherwise continue with the failure documented in the final report.

\subsection{Skill: \texttt{ds-cleaning}}
\label{app:skill-ds-cleaning}

\noindent\textbf{Purpose.} Two-mode data-quality sub-skill for either issue detection or issue-driven cleaning within the session sandbox.

\noindent\textbf{Modes.} (1)~\texttt{detect\_only}: detect missingness, type inconsistencies, duplicates, and simple outliers without producing cleaned data; (2)~\texttt{clean\_from\_issue\_report}: read the issue report and apply only focused cleaning actions.

\noindent\textbf{Workflow.} (1)~Write one stage script in the sandbox,
(2)~run it with \texttt{conda run -n muse --cwd ./sandbox/\$\{CLAUDE\_SESSION\_ID\} python <script.py>},
(3)~inspect the generated outputs,
(4)~write a concise markdown report,
(5)~mark the corresponding step complete.

\noindent\textbf{Outputs.} In detection mode: \texttt{data\_issue\_report.md}. In cleaning mode: \texttt{data\_cleaned.csv} and \texttt{data\_cleaning\_report.md}.

\noindent\textbf{Key Rules.} Operate only in the session sandbox; use fixed pipeline inputs rather than alternate datasets; prefer exactly one script execution per mode; keep runtime short; resolve paths to absolute locations inside the script; detection mode must not save cleaned data.

\subsection{Skill: \texttt{ds-eda}}
\label{app:skill-ds-eda}

\noindent\textbf{Purpose.} Moderate-depth exploratory data analysis sub-skill for producing a compact but informative set of univariate, bivariate, and correlation findings.

\noindent\textbf{Workflow.} (1)~Run in \texttt{mode=standard} with \texttt{max\_plots=8},
(2)~write one sandbox stage script,
(3)~generate a controlled set of 5--8 plots covering major distributions and relationships,
(4)~save figures as \texttt{.png} files in the sandbox,
(5)~inspect the outputs,
(6)~write a concise EDA report,
(7)~mark the EDA step complete.

\noindent\textbf{Outputs.} \texttt{eda\_report.md} and a bounded set of EDA figures such as \texttt{eda\_*.png}.

\noindent\textbf{Key Rules.} Use \texttt{data\_cleaned.csv} as the fixed input unless clearly invalid; never exceed 8 plots; avoid exhaustive or research-grade analysis; keep the workflow lightweight and path-safe under sandbox or resumed-session layouts.

\subsection{Skill: \texttt{ds-feature-engineering}}
\label{app:skill-ds-feature-engineering}

\noindent\textbf{Purpose.} Minimal feature-engineering sub-skill that prepares model-ready data through only necessary transformations.

\noindent\textbf{Workflow.} (1)~Run in \texttt{mode=minimal},
(2)~write one stage script in the sandbox,
(3)~apply required encoding,
(4)~apply scaling only when justified by the chosen baseline models,
(5)~add limited derived features only when supported by EDA findings,
(6)~save the engineered dataset,
(7)~write a concise feature-engineering report,
(8)~mark the step complete.

\noindent\textbf{Outputs.} \texttt{data\_engineered.csv} and \texttt{feature\_engineering\_report.md}.

\noindent\textbf{Key Rules.} No feature explosion; no unnecessary transforms; prefer compact runtime and outputs; use \texttt{data\_cleaned.csv} and \texttt{eda\_report.md} as fixed upstream inputs; keep all reads and writes inside the sandbox with resolved paths.

\subsection{Skill: \texttt{ds-modeling}}
\label{app:skill-ds-modeling}

\noindent\textbf{Purpose.} Fast baseline-modeling sub-skill for quick performance checks without heavy tuning.

\noindent\textbf{Workflow.} (1)~Run in \texttt{mode=baseline\_fast} with \texttt{max\_models=2} and \texttt{tuning=off},
(2)~write one sandbox modeling script,
(3)~use a single train/test split,
(4)~train only one or two baseline models,
(5)~evaluate with task-appropriate core metrics,
(6)~save the metrics file,
(7)~write a concise modeling report,
(8)~mark the step complete.

\noindent\textbf{Outputs.} \texttt{baseline\_metrics.json} and \texttt{baseline\_model\_report.md}.

\noindent\textbf{Key Rules.} No hyperparameter search; no long-running experiments; use \texttt{data\_engineered.csv}, \texttt{feature\_engineering\_report.md}, and \texttt{objective\_dataset\_report.md} as fixed inputs; keep execution deterministic and lightweight.

\subsection{Skill: \texttt{ds-progress-checker}}
\label{app:skill-progress-checker}

\noindent\textbf{Purpose.} Supervisory pipeline monitor that audits artifacts, determines completed and missing stages, and emits the next action for resume or recovery.

\noindent\textbf{Workflow.} (1)~Inspect only the session sandbox,
(2)~classify each stage as done, missing, or needing retry based on required artifacts,
(3)~respect the fixed dependency chain between stages,
(4)~determine whether the next action is to run a skill or finish,
(5)~write both markdown and JSON progress outputs,
(6)~emit a progress-check completion marker.

\noindent\textbf{Outputs.} \texttt{progress\_check\_report.md} and \texttt{progress\_check.json}, where the JSON contains completed stages, missing stages, invalid artifacts, and the next action specification.

\noindent\textbf{Key Rules.} Do not regenerate stage outputs; do not modify datasets or models; keep decisions minimal and deterministic; never emit \texttt{ask\_user}; default to continuing the lean pipeline when optional choices are unspecified.

\subsection{Skill: \texttt{step-complete}}
\label{app:skill-step-complete}

\noindent\textbf{Purpose.} Lightweight signaling utility that marks the completion of a meaningful pipeline phase for the UI or orchestration layer.

\noindent\textbf{Workflow.} (1)~When a major stage is fully finished,
(2)~run \texttt{echo "[STEP\_COMPLETE: <Step Name>]"},
(3)~use the plain human-readable step name only.

\noindent\textbf{Outputs.} A single completion marker line of the form \texttt{[STEP\_COMPLETE: <Step Name>]}.

\noindent\textbf{Key Rules.} Use this only for major self-contained phases, not after every tool call; do not include stage numbering or prefixes; the step name must be plain text such as \texttt{Data Cleaning} or \texttt{Baseline Modeling}.

\clearpage
\section{System Implementation}

{\tool}'s frontend is implemented using React~\cite{React18}, Vite~\cite{Vite}, 
and TypeScript~\cite{TypeScript}, with Tailwind CSS~\cite{TailwindCSS} 
for responsive styling. Outputs are rendered via React Markdown~\cite{ReactMarkdown}, 
React Syntax Highlighter~\cite{ReactSyntaxHighlighter}, 
and Recharts~\cite{Recharts}. The backend is built on FastAPI~\cite{FastAPI}, exposing RESTful APIs 
and a persistent WebSocket endpoint for real-time streaming of structured agent events. 
Agent execution is orchestrated using the Claude Agent SDK~\cite{ClaudeSDK} and its built-in tools, including \texttt{Read}, \texttt{Write}, \texttt{Edit}, 
\texttt{Bash}, \texttt{Glob}, \texttt{Grep}, and \texttt{Skill}. Semantic translations from L1--L5 are generated using Google Gemini APIs~\cite{GeminiAPI}, 
with explicit JSON schema constraints to enforce structured outputs. 
All LLM calls use \texttt{temperature=0}, following each model's default configuration, 
to ensure reproducibility.

\clearpage

\subsection{Prompt Design}

\begin{table*}[h!]
\small
\begin{tabular}{p{0.98\textwidth}}
\toprule
\textbf{CONTEXT}\\
You are given one semantic chunk from an AI-driven data workflow, including its execution log, script context, and report context.\\

\textbf{OBJECTIVE}\\
Produce a structured representation of this chunk that can support five levels of semantic detail in the user interface, from stage-level overview to implementation trace.\\

\textbf{GUIDELINES}\\
1. Use the chunk to infer what was accomplished in this workflow step.\\
2. Summarize the step in natural language before exposing detailed actions.\\
3. Group related operations into a small number of meaningful high-level tactics.\\
4. Preserve important implementation details, such as files, code actions, and outputs, for lower-level inspection.\\
5. Ground all content in the provided evidence and avoid unsupported inferences.\\

\textbf{INPUT}\\
- Step Name\\
- Execution Log\\
- Script Context\\
- Report Context\\

\textbf{OUTPUT}\\
Return structured JSON that supports:\\
- stage-level outcome,\\
- strategy summary,\\
- high-level plan,\\
- detailed operations, and\\
- underlying trace-level evidence.\\

\bottomrule
\caption{
Prompt template used for semantic chunk translation.
}
\label{tab:translation_prompt}
\end{tabular}
\end{table*}

\begin{table*}[h!]
\small
\begin{tabular}{p{0.98\textwidth}}
\toprule
\textbf{CONTEXT}\\
You are given one semantic step from an AI-driven data analysis workflow. Each step contains execution evidence, including the raw execution log, the current step's script context, and the report context.\\

\textbf{OBJECTIVE}\\
Determine whether the current step exhibits suspicious behavior that should be surfaced to the user as a warning.\\

\textbf{GUIDELINES}\\
1. Detect warnings only when they are supported by concrete evidence in the provided execution trace.\\
2. Use the script context as the primary source for detecting data-related issues.\\
3. Use the execution log together with the script context for detecting behavior-related issues.\\
4. Do not infer beyond the information provided in the input. If the evidence is weak or ambiguous, return no warning.\\
5. Be conservative: only flag issues that could plausibly affect the correctness, reliability, or interpretability of the step outcome.\\
6. Each warning must be a short sentence that names both the risk and the supporting evidence.\\
7. If warnings are detected, generate short and actionable revision suggestions that directly address the identified issues.\\

\textbf{INPUT}\\

- Warning Rules: refer to Table~\ref{tab:detailed_warning_heuristics}.\\
- Step Name: the name of the current workflow step.\\
- Execution Log: the raw log of actions, commands, and outputs for the current step.\\
- Script Context: the code written or executed in the current step.\\
- Report Context: any report or summary text associated with the step.\\

\textbf{WARNING TYPES}\\
Warnings may capture issues such as suspicious data processing, destructive or unjustified transformations, missing verification, mismatches between stated reasoning and executed actions, incorrect or incomplete outputs, and premature termination.\\

\textbf{OUTPUT}\\
Your output must be valid JSON in the following format:\\
\{\\
\qquad "summary": "<one-sentence description of what the step did>",\\
\qquad "data\_level\_warnings": ["<short data-related warning>"],\\
\qquad "agent\_level\_warnings": ["<short behavior-related warning>"],\\
\qquad "revision\_suggestions": ["<short actionable suggestion 1>", "<short actionable suggestion 2>", "<short actionable suggestion 3>"]\\
\}\\
If no warning is supported by clear evidence, return empty lists for both warning fields and for \texttt{revision\_suggestions}.\\

\bottomrule
\caption{
The prompt used for warning detection in \tool.
}
\label{tab:warning_prompt}
\end{tabular}
\end{table*}

\begin{table*}[h!]
\small
\begin{tabular}{p{0.98\textwidth}}
\toprule
\textbf{CONTEXT}\\
You are a QA assistant for a data science pipeline application. A separate execution agent runs the pipeline and writes artifacts into a session workspace. Your role is to answer the user's questions about the current pipeline session without modifying files or rerunning work.\\

\textbf{OBJECTIVE}\\
Answer the user's question about the selected workflow step or artifact from the available session context.\\

\textbf{GUIDELINES}\\
1. Answer based only on the current session context and any read-only inspection you perform.\\
2. Prefer progress artifacts and stage reports when they conflict with setup summaries.\\
3. Never invent stage names, stage numbers, or unofficial sub-stages.\\
4. Do not invent facts.\\
5. Do not modify files or take over execution from the pipeline worker.\\
6. Keep file reads bounded and focused on the session workspace.\\
7. If the session context is incomplete, say what is missing instead of guessing.\\

\textbf{INPUT}\\
- User Question\\
- Session Context\\
- Available Session Files\\
- Candidate Relevant Files\\
- Recent Internal User--Agent Dialog\\
- Optional Focused Warning Context\\

\textbf{OUTPUT}\\
Return a concise natural-language answer grounded in the current session context.\\
If the question asks about progress, answer in terms of completed stage, current stage, and next stage when possible.\\

\bottomrule
\caption{
Prompt template used by \tool for contextualized step referencing.
}
\label{tab:contextualized_referencing_prompt}
\end{tabular}
\end{table*}

\begin{table*}[h!]
\small
\begin{tabular}{p{0.98\textwidth}}
\toprule
\textbf{CONTEXT}\\
You are continuing an existing workflow run. A previously executed step has been flagged as suspicious and should be revised in context.\\

\textbf{OBJECTIVE}\\
Translate the user's high-level revision intent into a contextualized handoff request for the underlying DS agent.\\

\textbf{GUIDELINES}\\
1. Treat the selected step as the restart point for revision.\\
2. Use the warning context to understand why the step is being revised.\\
3. Apply the user's revision intent to the target step.\\
4. Continue all downstream steps from the revised step onward.\\
5. Overwrite downstream artifacts that are no longer valid after the revision.\\
6. Keep all execution and outputs within the current session sandbox.\\

\textbf{INPUT}\\
- Step Name: the workflow step to revise.\\
- Step Order: the position of the step in the workflow.\\
- Warning Context: the detected issues associated with the step.\\
- User Revision Plan: the user's requested fix or modification.\\

\bottomrule
\caption{
Template used to construct contextualized revision handoff requests.
}
\label{tab:revise_handoff_template}
\end{tabular}
\end{table*}

\begin{table*}[h!]
\small
\begin{tabular}{p{0.98\textwidth}}
\toprule
\textbf{CONTEXT}\\
You are helping a user verify a file or selected output from an AI-driven data workflow. The user has highlighted a specific file or text span and wants to check whether the result is correct or well supported.\\

\textbf{OBJECTIVE}\\
Generate a small set of concrete verification options grounded in the selected file or content.\\

\textbf{GUIDELINES}\\
1. Base each verification option on the selected file or text span.\\
2. Focus on actionable checks related to correctness, data quality, consistency, or workflow integrity.\\
3. Do not propose generic or overly broad checks.\\
4. Keep each option short and easy for the user to choose from.\\
5. Generate exactly 3--5 verification options.\\

\textbf{INPUT}\\
- File Path: the file selected by the user.\\
- Selection: the highlighted content or excerpt to verify.\\
- Session Context: the current workflow session and its associated artifacts.\\

\textbf{OUTPUT}\\
Return 3--5 specific verification options that the user might want to perform on the selected file or output.\\

\bottomrule
\caption{
Prompt template used to generate scaffolded verification options.
}
\label{tab:verification_prompt}
\end{tabular}
\end{table*}

\begin{table*}[h!]
\small
\begin{tabular}{p{0.98\textwidth}}
\toprule
\textbf{CONTEXT}\\
You are verifying a user-selected output from an existing workflow session. The workflow artifacts are stored in a session-scoped sandbox, and your role is to inspect them without modifying canonical pipeline outputs.\\

\textbf{OBJECTIVE}\\
Carry out the selected verification request using the current session context and determine whether the result is supported by the available evidence.\\

\textbf{GUIDELINES}\\
1. Inspect only files and artifacts relevant to the selected output.\\
2. Use the current session sandbox as the primary evidence source.\\
3. Do not modify official pipeline outputs or reports.\\
4. When simple inspection is insufficient, you may create lightweight verification artifacts inside the session sandbox.\\
5. If verification artifacts are created, keep them advisory and separate from canonical workflow outputs.\\
6. Clearly state whether the verification passed, failed, or remains inconclusive.\\
7. Cite the relevant files used as evidence.\\

\textbf{INPUT}\\
- User Verification Request\\
- Selected Output or Artifact Reference\\
- Session Context\\
- Relevant Files\\

\textbf{OUTPUT}\\
Return a concise verification result grounded in the inspected workflow artifacts. If needed, create lightweight verification artifacts such as \texttt{muse\_verify\_<topic>.py}, \texttt{muse\_verify\_<topic>.md}, or \texttt{muse\_verify\_<topic>.json}.\\

\bottomrule
\caption{
Prompt template used for verification execution.
}
\label{tab:verification_execution_prompt}
\end{tabular}
\end{table*}

\begin{table}[t]
\centering

\renewcommand{\arraystretch}{1.15}
\setlength{\tabcolsep}{6pt}
\begin{tabular}{p{\linewidth}}
\toprule
\textbf{Detailed Warning Heuristics} \\
\midrule

\multicolumn{1}{l}{\textbf{Data-level Warnings}} \\
\midrule

\textbf{Aggressive Data Removal.}
This warning is triggered when the script removes rows or columns too broadly, destructively, or without sufficient justification, such that important information may be discarded and the remaining dataset may become biased or unrepresentative. \\

\textbf{Overwriting Raw Data.}
This warning applies when transformed or cleaned outputs are written back to the original data source, so that the raw input is no longer preserved for reproducibility, auditing, or recovery. \\

\textbf{Hard-coded Thresholds or Magic Numbers.}
This warning captures the use of fixed cutoffs, constants, binning boundaries, or manually chosen numeric rules that are embedded in code without visible empirical, domain, or methodological justification. \\

\textbf{Suspicious Joins or Merges.}
This warning is raised when datasets are merged in ways that may duplicate observations, drop records, or misalign entities, especially when join keys, join types, or cardinality checks are not clearly validated. \\

\textbf{Lost Categorical Semantics.}
This warning occurs when categorical variables are converted into in-memory categorical codes or labels, and then saved to flat outputs such as CSV without preserving the metadata needed to recover their original semantic meaning. \\

\textbf{Missing or Misleading Evaluation.}
This warning indicates that model performance is reported without a clear assessment on validation, holdout, or test data, or that the reported results are misleading because they rely only on training performance. \\

\textbf{Single-Split Evaluation.}
This warning refers to model assessment based on only one train/test split or one-off holdout evaluation, without cross-validation, repeated resampling, or other procedures that would provide a more robust estimate of performance. \\

\textbf{Unjustified Transformations.}
This warning is used when the script applies transformations, encodings, filters, or feature manipulations whose rationale, validity, or expected analytical benefit is not evident from the code. \\

\textbf{Inefficient Data Processing.}
This warning captures unnecessarily slow, wasteful, or poorly scalable processing choices for the workload at hand, such as avoidable row-wise operations, repeated full-data passes, or excessive intermediate materialization. \\

\textbf{Redundant Data Operations.}
This warning applies when the script repeatedly loads, copies, transforms, or writes the same data in ways that do not contribute clear analytical value and may introduce unnecessary complexity or inconsistency. \\

\textbf{Unverified Schema Assumptions.}
This warning is raised when the script directly assumes the existence of specific columns, field names, or schema structure without visibly checking that the required inputs are actually present. \\

\textbf{Missing Data Not Inspected.}
This warning occurs when the script proceeds to analysis, visualization, feature engineering, or summary statistics without any visible inspection of missingness, such as null counts, missing-value summaries, or explicit checks. \\

\textbf{Invalid Categorical Interaction Encoding.}
This warning refers to interaction features constructed by numerically combining categorical codes, such as multiplying encoded categories together, which can impose false ordinal structure or create arbitrary collisions. \\

\midrule
\multicolumn{1}{l}{\textbf{Agent-level Warnings}} \\
\midrule

\textbf{Reasoning--Action Mismatch.}
This warning is triggered when the agent's stated plan or claimed intention does not match the actions actually taken in the log, indicating a gap between reasoning and execution. \\

\textbf{Task Derailment.}
This warning applies when the agent drifts away from the main objective of the step and spends effort on tangential or irrelevant actions that do not materially advance the assigned task. \\

\textbf{Insufficient Verification.}
This warning is raised when the agent claims completion or success without adequately validating outputs, checking metrics, inspecting artifacts, or confirming that the produced results are correct and usable. \\

\textbf{Unverified Assumptions.}
This warning captures cases where the agent proceeds based on assumptions about files, data, schema, environment, or expected behavior without first confirming that those assumptions are true. \\

\textbf{Role Misalignment.}
This warning occurs when the agent performs work that does not match the intended responsibility of the current stage, such as skipping the expected step objective or acting outside the designated role. \\

\textbf{Task Specification Violation.}
This warning indicates that the agent fails to follow explicit instructions, constraints, or deliverable requirements defined for the step, even if some useful work is still produced. \\

\textbf{Premature Termination.}
This warning refers to cases where the agent stops too early, exits before producing the expected artifacts, or ends the step before key actions or validations have been completed. \\
\bottomrule

\end{tabular}
\caption{Detailed descriptions of warning heuristics in \tool.}
\label{tab:detailed_warning_heuristics}
\end{table}

\begin{figure}[h]
\vspace{-1pt}
    \centering
    \includegraphics[width=0.6\linewidth]{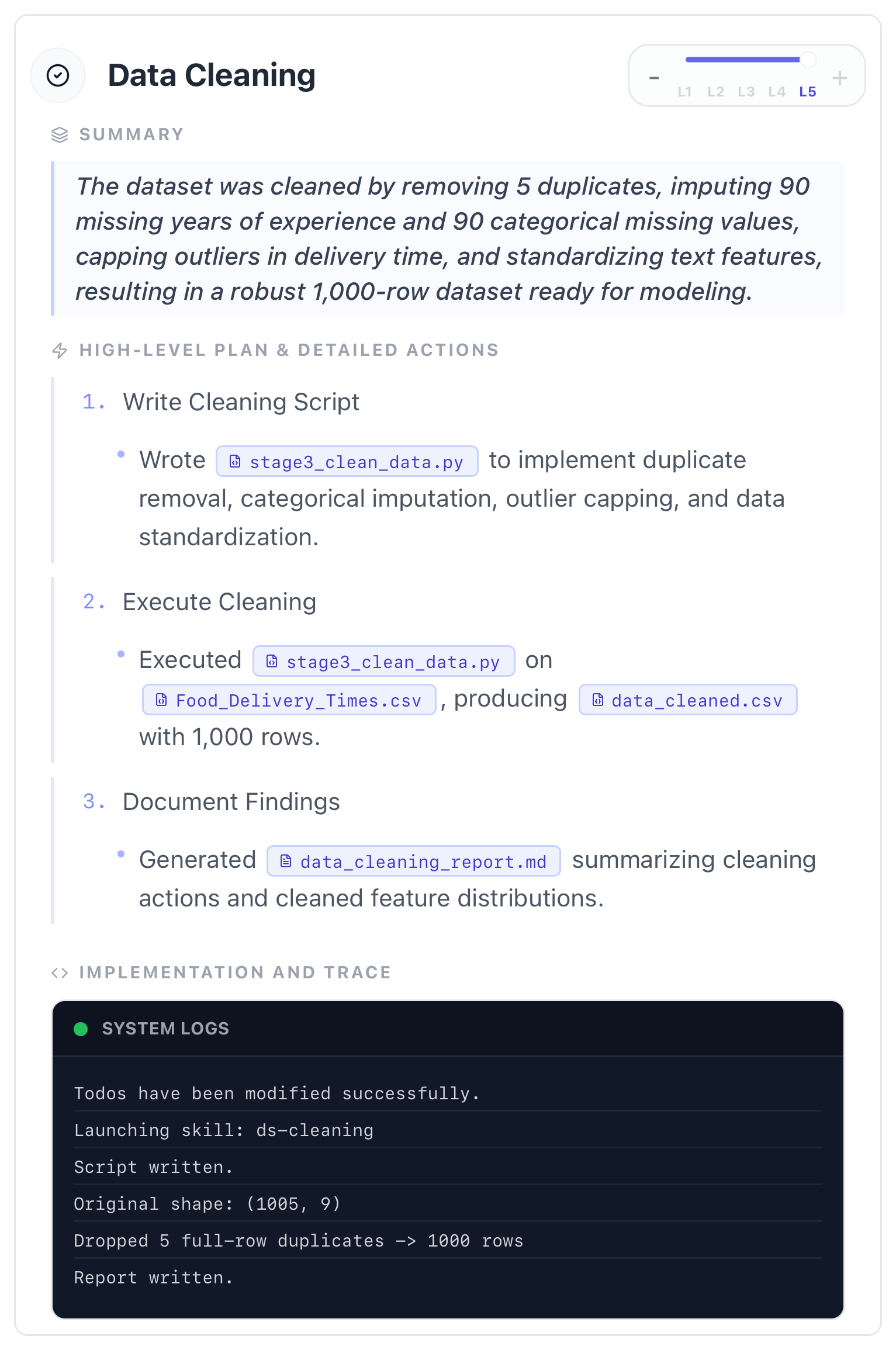}

    \caption{An example of the multi-level semantic representation for a data cleaning stage.}
    \label{fig:data_cleaning_card}
    
\end{figure}
\end{document}